\documentclass[aps,prd,preprintnumbers,superscriptaddress,floatfix,amssymb,amsfonts,twocolumn]{revtex4}  
\usepackage{hyperref}
\usepackage[utf8]{inputenc}\DeclareUnicodeCharacter{2212}{-}
\usepackage{amsmath}
\usepackage[normalem]{ulem}
\usepackage{import}
\usepackage{graphicx}
\usepackage{amssymb}
\usepackage{multirow}
\usepackage{mathbbol}
\usepackage{color}
\usepackage{makecell}

\newcommand{\brackets}[1]{\langle #1 \rangle}

\newcommand{\epow}[1]{\mathrm{e}^{#1}}

\newcommand{\gammafive}{\gamma_{5}}

\newcommand{\fermi}{\,\mathrm{fm}}
\newcommand{\gev}{\,\mathrm{GeV}}
\newcommand{\mev}{\,\mathrm{MeV}}

\newcommand{\dbar}{\bar{d}}

\newcommand{\balign}{\begin{align}}
\newcommand{\ealign}{\end{align}}
\newcommand{\beq}{\begin{equation}}
\newcommand{\eeq}{\end{equation}}
\newcommand{\balignat}[1]{\begin{alignat}{#1}}
\newcommand{\ealignat}{\end{alignat}}

\newcommand{\bfig}{\begin{figure}}
\newcommand{\efig}{\end{figure}}

\newcommand{\bc}{\begin{center}}
\newcommand{\ec}{\end{center}}

\newcommand{\btab}{\begin{table}}
\newcommand{\etab}{\end{table}}

\newcommand{\bcom}{}

\newcommand{\bitem}{\begin{itemize}}
\newcommand{\eitem}{\end{itemize}}

\newcommand{\benum}{\begin{enumerate}}
\newcommand{\eenum}{\end{enumerate}}

\newcommand{\pvec}{\vec{p}}
\newcommand{\Pvec}{\vec{P}}

\newcommand{\xvec}{\vec{x}}

\newcommand{\yvec}{\vec{y}}

\newcommand{\zvec}{\vec{z}}

\newcommand{\dvec}{\vec{d}}

\newcommand{\refeq}[1]{(\ref{#1})}

\newcommand{\qcmf}{q_{\mathrm{cmf}}}

\newcommand{\Nbar}{\bar{N}}

\begin{document}

\title{Elastic Nucleon-Pion scattering amplitudes in the $\Delta$ channel at physical pion mass from Lattice QCD}

\author{Constantia Alexandrou} \affiliation{Department of Physics,
  University of Cyprus, P.O. Box 20537, 1678 Nicosia, Cyprus}
\affiliation{Computation-based Science and Technology Research Center,
  The Cyprus Institute, 20 Kavafi Str., Nicosia 2121, Cyprus}

\author{Simone Bacchio} \affiliation{Computation-based Science and
  Technology Research Center, The Cyprus Institute, 20 Kavafi Str.,
  Nicosia 2121, Cyprus}

\author{Giannis Koutsou} \affiliation{Computation-based Science and
  Technology Research Center, The Cyprus Institute, 20 Kavafi Str.,
  Nicosia 2121, Cyprus}

\author{Theodoros Leontiou} \affiliation{Frederick Research Center, 7
  Filokyprou St Palouriotissa, 1036, Nicosia, Cyprus}

\author{Srijit Paul}  \affiliation{
  Higgs Center for Theoretical Physics, The University of Edinburgh,
  Peter Guthrie Tait Road
  Edinburgh, EH9 3FD, UK }

\author{Marcus Petschlies}  \affiliation{HISKP (Theory), Rheinische
  Friedrich-Wilhelms-Universit{\"a}t Bonn, Nu{\ss}allee 14-16, 53115
  Bonn, Germany}

\author{Ferenc Pittler} \affiliation{Computation-based Science and
  Technology Research Center, The Cyprus Institute, 20 Kavafi Str.,
  Nicosia 2121, Cyprus}

\date{\today}

\begin{abstract} 
  We present an investigation of pion-nucleon elastic scattering in
  the $I\,(J^P) = \frac{3}{2}\,(\frac{3}{2}^+)$ channel using lattice
  QCD with degenerate up and down, strange and charm quarks with
  masses tuned to their physical values.  We use an ensemble of
  twisted mass fermions with box size $L = 5.1\,\mathrm{fm}$ and
  lattice spacing $a = 0.08\,\mathrm{fm}$ and we consider the $\pi N$
  system in rest and moving frames up to total momentum $\vec{P}^2 =
  3\,(2\pi/L)^2$ = 0.17~GeV$^2$. We take into account the finite
  volume symmetries and $S$- and $P$-wave mixing, and use the
  L\"uscher formalism to simultaneously constrain the $J = 1/2,\,\ell
  = 0$ and $J = 3/2,\,\ell = 1$ scattering amplitudes. We estimate the
  $\Delta$ resonance pole in the $P$-wave channel as well as the
  $S$-wave isospin-3/2 scattering length.
\end{abstract}

\maketitle

\section{Introduction}
The precise treatment of nucleon resonances is still a formidable
task in lattice QCD. While L\"uscher's method
\cite{Luscher:1986pf,Luscher:1990ux,Luscher:1991cf} is the established
theoretical basis for connecting observed lattice QCD spectra to
scattering amplitudes, thus allowing the investigation of properties
of bound states and resonances from first principles, in practice the
study of meson-baryon, two-hadron states remains challenging. This is 
especially so when using simulations with physical values of the light
quark mass, which carry increased statistical errors.

Nevertheless, the \textit{ab initio} computation of low-energy elastic
pion-nucleon ($\pi N$) scattering from lattice QCD is essential for
the study of nucleon interactions, and any such treatment necessarily
starts with the lowest-lying meson-baryon resonance, namely the
$I\,(J^P) = \frac{3}{2}\,(\frac{3}{2})^+$ $\Delta$ $P$-wave resonance.
The $\Delta$ resonance governs nucleon-pion, nucleon-photon, and
nucleon-neutrino scattering as the dominant channel. Within the
lattice QCD formalism, excited state contributions from pion-nucleon
scattering states dominate the spectrum in nucleon form factor
calculations in a finite volume for gauge ensembles generated with
close to physical pion mass~\cite{Bar:2019zkx, Bar:2019igf}.

The interaction of nucleon and pion has been studied by various
approaches in lattice QCD in the past, using gauge ensembles generated
with heavier-than-physical pion masses. Nucleon-pion scattering
amplitudes and the $\Delta$ in particular have been the subject of
Refs.~\cite{Bulava:2022vpq, Silvi:2021uya, Pittler:2021bqw,
  Andersen:2017una, Lang:2016hnn, Verduci:2014btc, Lang:2012db,
  Mohler:2012nh} using the L\"uscher method.
Refs.~\cite{Alexandrou:2013ata, Alexandrou:2015hxa} used an
alternative method based on Refs.~\cite{McNeile:2002az,
  McNeile:2002fh}.  There are also studies using ensembles generated
with quark masses at which the $\Delta $ is a stable
state~\cite{Engel:2013ig, Dudek:2012ag, Edwards:2011jj, Engel:2010my,
  Gattringer:2008vj, Basak:2007kj}.

With this work, we extend the lattice calculation of the $\Delta$ from
heavier-than-physical to physical pion mass, and explore the
application of the L\"uscher method to the $\pi N - \Delta$ channel at
the physical point. We estimate the $\Delta$ resonance pole in the
$P$-wave channel as well as the $S$-wave isospin-3/2 scattering
length, which experimentally enters the evaluation of the pion-nucleon
$\sigma$-term using the Roy-Steiner-equations. For this first physical
point calculation, we use a single ensemble with two degenerate light
quarks, strange, and charm quarks ($N_f=2+1+1$).

The paper is organized as follows: In
Sec.~\ref{sec:scattering-formalism}, we describe our application of
the L\"uscher method, with further details on the implementation of
the lattice spectroscopy in
Sec.~\ref{sec:methods}. Secs.~\ref{sec:spectrum_analysis}
and~\ref{sec:scattering_analysis} present our analysis of the lattice
data for the $\pi N-\Delta$ spectrum, and subsequent fits to the
finite-volume quantization condition. In addition, we use the
threshold expansion of the L\"uscher quantization condition to
determine the $S$-wave scattering length. A discussion of the results
and concluding remarks are given in Sec.~\ref{sec:conclusions}. In
appendices~\ref{appendix:stability-plots},~\ref{appendix:tables-results},
and~\ref{appendix:optimization} we include extended figures and tables
of our results, as well as some additional details of our analysis.

\section{Lattice QCD formalism for pion-nucleon scattering}
\label{sec:scattering-formalism}

We consider elastic scattering of two particles of non-equal mass and
with spin in a hypercubic box of spatial extent $L$. The constraints
on infinite-volume scattering amplitudes from the finite volume
lattice spectrum for this setup are given in L\"uscher's original work
\cite{Luscher:1986pf,Luscher:1990ux,Luscher:1991cf} and a series of
extensions, to moving frames~\cite{Kim:2005gf,Rummukainen:1995vs}, to
non-degenerate particles~\cite{Leskovec:2012gb} and to particles with
spin~\cite{Morningstar:2017spu,Briceno:2014oea,Briceno:2013lba,Gockeler:2012yj}.
While the formalism for multiple coupled two-particle decay channels
is also known~\cite{Hansen:2012tf,He:2005ey}, for the purposes of this
work we perform a single decay channel analysis, since the expected
branching fraction $\Delta \to \pi N$ is almost $100 \%$ making this
channel a prototype application for lattice QCD.

The finite-volume method for three particles has been further
developed for spin-less particles
\cite{Briceno:2019muc,Jackura:2020bsk,Hansen:2020zhy,Hansen:2021ofl}. For
the $N\pi\pi$ system a formalism is not yet available, and we thus
only consider the $\pi N$ interaction here. We will define an upper
limit to the spectrum entering our elastic scattering amplitude fits,
such that the impact of the three-particle scattering is negligible.

For the pion-nucleon system, even and odd partial waves mix. Moreover,
the spin of the nucleon couples to the orbital angular momentum, such
that in the finite-volume analysis the most relevant partial wave
amplitudes are $J = 1/2,$ containing $S$- and $P$-wave, and $J = 3/2$,
with $P$- and $D$-wave.  The $J=3/2$ $P$-wave amplitude corresponds to the
$\Delta$ resonance and is expected to dominate, with a sub-leading
contribution from $J=1/2$ $S$-wave, while amplitudes from the
corresponding higher $\ell$ values are suppressed
by angular momentum barrier.

To the extent described above, the L\"uscher quantization conditions
have been given in detail in
Refs.~\cite{Bernard:2008ax,Gockeler:2012yj} with the master equation given by 
\begin{align}
  \mathrm{det}\left( 
  M^{\Pvec,\Lambda}_{J\ell n; J' \ell' n'} - \delta_{J J'}\,\delta_{\ell \ell' }\,\delta_{n n'} \, \cot(\delta_{J \ell})
  \right) &= 0,
  \label{eq:lqc}
\end{align}
where $M^{\Pvec,\Lambda}$ denotes the reduced L\"uscher finite volume
matrix for a reference frame with total momentum $\Pvec$ of the $\pi N$ system, and for irreducible representation (irrep) $\Lambda$ of
the little group $LG(\Pvec) \subseteq O^D_h$.  Here we use the reduced
L\"uscher matrices and quantization conditions from
Refs.~\cite{Bernard:2008ax,Gockeler:2012yj}.  The determinant is taken
in the linear space of total angular momentum $J, J'$, orbital angular
momentum $\ell, \ell'$ and multiplicity (or occurrence) of the irrep
$\Lambda$ $n, n'$.

For ensembles  with physical value of the pion mass, the thresholds
for elastic nucleon-pion scattering are given by $E_{2-\mathrm{thr}} =
m_{N} + m_{\pi} \approx 1080 \mev$ and the three-particle threshold
$E_{3-\mathrm{thr}} = m_{N} + 2 m_{\pi} \approx 1220 \mev$. The
correspondingly narrow window in the center of mass energy for elastic $\pi N$ scattering does not cover the expected resonance region and puts
prohibitive cuts on the usable finite-volume lattice
spectrum. However, based on experimental observations, the $\pi N$
scattering amplitude in the $I = 3/2$ channel is vastly dominated by
the elastic two-particle interaction up to $E_{3-\mathrm{thr}}' =
m_{\Delta} + m_{\pi}$ as a proxy three-particle threshold, where the
$\Delta$ and the pion can propagate on-shell \cite{Ronchen:2012eg}. We
thus consider lattice energy levels up to $\sqrt{s} \lesssim
E_{3-\mathrm{thr}}' \approx 1360 \mev$.

The partial wave amplitudes are parameterized via the associated phase
shifts. In particular, for the considered leading phase shifts
$\delta_{J\ell}$, we employ the analytic Breit-Wigner form for the
resonant $\Delta$ channel and the leading order effective range
expansion with isospin $I=3/2$ $S$-wave scattering length $a_0$.
\begin{align}
  \cot\left( \delta_{\frac{1}{2}0}(s) \right) &= a_0\,\qcmf \,,
  \label{eq:delta-s-wave} \\
  \tan\left( \delta_{\frac{3}{2}1}(s) \right) &= \frac{\sqrt{s}\,\Gamma\left( \Gamma_R,\,M_R, \,s \right)}{M_R^2 - s} \,.
  \label{eq:delta-p-wave}
\end{align}
The center of mass momentum ($\qcmf$) of the $\pi N$ system is given by
\begin{align}
  \qcmf^2(s) &= \frac{\left( s - M_{N}^2 - M_{\pi}^2 \right)^2 - 4\,M_N^2\,M_{\pi}^2 }{4s} \,,
  \label{eq:qcmf}
\end{align}
and the resonance decay width has invariant mass dependence parameterized by
\begin{align}
  \Gamma\left( \Gamma_R,\,M_R,\,s \right) &= \Gamma_R\,\left( \frac{\qcmf(s)}{\qcmf(M_R^2)} \right)^3\,\frac{M_R^2}{s} \,.
  \label{eq:gamma-width}
\end{align}

\section{Correlation matrix construction}
\label{sec:methods}
\subsection{Interpolating operators}
To constrain pion-nucleon scattering amplitudes using the L\"uscher
method, we determine the low-lying finite-volume energy spectrum of
the lattice Hamiltonian with isospin $I=3/2$. The operator basis for
the correlation matrices of two-point functions is constructed from
single- and two-hadron timeslice interpolating operators. We consider
the case of maximal isospin, i.e. $I_3 = +3/2$, meaning we use the
$\Delta^{++}$, and the proton and charged pion ($N^{+}$ and $\pi^{+}$).

The single-hadron, quark model $\Delta$-type interpolating operator
reads
\begin{align}
  \left(O_{\Delta^{++}}\right)_{\alpha,k} (\Pvec; t) &= 
  \nonumber\\
    \sum_{\xvec}\,\varepsilon_{abc} & \,u^a_{\alpha}(\xvec,t)\, \left[
    u^{b}(\xvec,t)^\top\,\mathcal{C}\gamma_k\,u^c(\xvec,t) \right]\,e^{i\Pvec\xvec}
    \,,
    \nonumber\\
    k &= x,y,z \,.
  \label{eq:Delta-op}
\end{align}
The two-hadron interpolators are generated by products of nucleon
and pion interpolators given by
\begin{align}
  \left(O_{N^+\pi^+}\right)_{\alpha} (\pvec_N,\pvec_\pi; t)
  &= N^+_\alpha(\pvec_N ; t)\, \pi^+(\pvec_\pi;t),
  \label{eq:Npi-op}
\end{align}
with
\begin{align}
  N^+_{\alpha}(\pvec; t) 
    &= \sum\limits_{\xvec}\,\varepsilon_{abc}\,u^a_{\alpha}(\xvec,t)\, \left[ u^b(\xvec,t)^\top\,\mathcal{C}\gamma_5\,d^c(\xvec,t) \right]\,e^{i\pvec\xvec}
  \label{eq:N-op} \\
  \pi^+(\pvec,t) &= \sum\limits_{\xvec}\,\dbar(\xvec,t)\,\gammafive\,u(\xvec,t) \,e^{i\pvec\xvec},
  \label{eq:pi-op}
\end{align}
where $u$ and $d$ are up- and down-quark spinor fields,
$\mathcal{C}=\gamma_4\gamma_2$ is the charge-conjugation matrix, and
$\alpha$ denotes the spinor index. The total momentum of such interpolating fields is $\Pvec = \pvec_N + \pvec_{\pi}$.

\begin{widetext}
  
\begin{table}
  \centering
  \caption{Momentum frames (first column), lattice rotation symmetry
    groups (second column) and irreducible representations $\Lambda$
    (third, fourth, and fifth columns) with their subduced angular
    momentum content $J$, relevant for the pion ($J^P = 0^-$), 
    the nucleon ($J^P = 1/2^{+}$), and the $\Delta$-baryon ($J^P = 3/2^{+}$).  The
    naming of irreps follows Ref.~\cite{Morningstar:2013bda}.
  }
  \label{tab:irreps}
  \begin{tabular}{lcccc}
    \hline\hline
    \multirow{2}{*}{$\Pvec\,\left[ 2\pi/L \right]$} & \multirow{2}{*}{$LG(\Pvec)$} & \multicolumn{3}{c}{$I$} \\
                                   &             & $1$                       & $1/2$                      & $3/2$ \\
    \hline
    $(0,0,0)$                      & $O^D_h$     & $A_{1u}(0^{-},\,4^{-},\,\cdots)$    & $G_{1g}(1/2^{+},\,7/2^{+},\cdots)$      & $H_{g}(3/2^{+},\,5/2^{+},\cdots)$ \\
    $(0,0,1)$                      & $C^D_{4v}$  & $A_1(0,\,1,\,\cdots)$              & $G_{1}(1/2,\,3/2,\cdots)$              & \makecell{$ G_1(1/2,\,3/2,\cdots )$ \\ $ G_2(3/2,\,5/2,\cdots)$} \\
    $(0,1,1)$                      & $C^D_{2v}$  & $A_1(0,\,1,\,\cdots)$              & $G(1/2,\,3/2,\cdots)$                  & $G(1/2,\,3/2,\cdots)$ \\
    $(1,1,1)$                      & $C^D_{3v}$  & $A_1(0,\,1,\,\cdots)$              & $G(1/2,\,3/2,\cdots)$                  & \makecell{$ G(1/2,\,3/2,\cdots)$ \\ $ F_1(3/2,\,5/2,\cdots)$ \\ $ F_2(3/2,\,5/2,\cdots)$ }  \\
    \hline
    \hline
  \end{tabular}
\end{table}

\end{widetext}

The above operators are projected to irreducible representations of
the lattice rotation symmetry groups ($O^D_h$ for $\Pvec = 0$ or
little group $LG(\Pvec)$) for the rest and moving frames. The values
of the total momenta $\Pvec$ and lattice irreps used are given in
Table~\ref{tab:irreps}, where we also indicate the subduced angular
momenta.

The projection to irrep $\Lambda$, row $r$ for occurrence $n$, follows
from the group theory master formula, namely for the single hadron
operators $O_{\Delta}$ and the $\pi N$ operators $O_{\pi N}$ we use
\begin{widetext}
\begin{align}
  O^{\Lambda,r,n}_{\Delta;\,\alpha,k}(\Pvec)^\dagger &= \frac{\mathrm{dim}(\Lambda)}{\#LG(\Pvec)}\,\sum\limits_{G \, \in \,LG(\Pvec)}\,\Gamma^{\Lambda}(G)_{r,r'}^*\,
  U^{(1)}(G)_{k',k}\,U^{(1/2\oplus 1/2)}(G)_{\alpha',\alpha}\,\Delta_{\alpha',k}(\Pvec)^{\dagger} \,,\nonumber\\
  O^{\Lambda,r,n}_{N \pi;\,\alpha}(\Pvec,\pvec)^\dagger &= \frac{\mathrm{dim}(\Lambda)}{\#LG(\Pvec)}\,\sum\limits_{G \, \in \, LG(\Pvec)}\,\Gamma^{\Lambda}(G)_{r,r'}^*\,
  U^{(1/2\oplus 1/2)}(G)_{\alpha',\alpha}\,\Nbar_{\alpha'}\left(R\pvec\right)\,\pi(\Pvec-R\pvec)^{\dagger} \,,
  \label{eq:projection}
\end{align}
\end{widetext}
where the group element $G\, \in\, LG(\Pvec)$ means either a proper
rotation $R$ or a rotation-inversion $IR$, such that $R\Pvec = -\Pvec$
and concatenation with spatial inversion $I$ leaves invariant 
$\Pvec$.  Analogously, $U^{(J)}(G)$ denotes the $SU(2)$ spin-$J$
representation matrix of the proper rotation or rotation-inversion
group element.  The rotation matrix $U^{(1)}$ acting on the momentum
vector $\pvec$ in Cartesian basis, is denoted as $R$ for simplicity.
Moreover, $\Delta_{\alpha,k}$ and $N_{\alpha}$ are Dirac four-spinors
and the rotation matrix for the four-component spinors is denoted as
$U^{(1/2\oplus 1/2)}$. The space inversion operation $(\xvec,\,t) \to
(-\xvec,\,t)$ is represented by $\pi^+(\xvec,t) \to
-\pi^{+}(-\xvec,t)$ for the pseudoscalar pion field, and by
$\Delta_{\alpha,k}(\xvec,t) \to
(\gamma_4)_{\alpha\beta}\,\Delta_{\beta,k}(-\xvec,t) $ for the
four-component nucleon and $\Delta$ spinors.

The constructed set of irrep-projected operators $\left\{
O^{\Lambda,r,n}_{\Delta;\,\alpha,k}(\Pvec),\,
O^{\Lambda,r,n}_{N\pi;\,\alpha}(\Pvec,\pvec)\right\}$ is still linearly
dependent and we extract a basis set by the Gram-Schmidt
procedure. The orthogonalization is done with respect to the tensor
components $\alpha,\,k$ and momentum vector $\pvec$.

\subsection{Equivalent moving frames}
The explicit subduction coefficients for the operators, which are
obtained based on the convention in Ref.~\cite{Morningstar:2013bda},
pertain to the reference moving frames with total momentum
$\Pvec/(2\pi/L) = (0,0,1),\,(0,1,1),\,(1,1,1)$.  We include lattice
correlation functions also from all other equivalent moving frames of
the same orbit as $\Pvec$ under discrete rotations.  Little groups
$LG(\Pvec')$ and $LG(\Pvec)$ with $\Pvec^{2} = \Pvec^{'2}$ are
equivalent for the values considered here, namely $\left\{ \Pvec^2 \le
3\,(2\pi/L)^2 \right\}$, and we select one group element $R_{\Pvec'}
\,\in\,O^D_h$ per case, which relates $\Pvec' = R_{\Pvec'}\,\Pvec$.

\begin{table}
   \caption{Rotation group elements ($R$) to map moving frames
     $\vec{P}'$ to their reference directions.  The first set of two
     columns correspond to reference moving frame
     $\Pvec/(2\pi/L)=(0,0,1)$, the second to $(0,1,1)$, and the third
     set to $(1,1,1)$.  The superscript $i$ denotes the inverse group
     element.}
    \label{tab:mf-rots}  
    \begin{tabular}{cc | cc | cc}
      \hline\hline
     $\Pvec'/(2\pi/L)$  & R & $\Pvec'/(2\pi/L)$  & R &  $\Pvec'/(2\pi/L)$  & R \\\hline
      $(\phantom{-}0,\phantom{-}1,\phantom{-}0)$  &       $C_{4x}^i$ &  $(\phantom{-}1,-1,\phantom{-}0)$  &       $C_{2c}$   &  $(\phantom{-}1,\phantom{-}1,-1)$  &  $C_{4y}$   \\
      $(\phantom{-}1,\phantom{-}0,\phantom{-}0)$  &       $C_{4y}$   &  $(\phantom{-}1,\phantom{-}0,\phantom{-}1)$   &       $C_{4z}^i$ &	 $(\phantom{-}1,-1,\phantom{-}1)$  &  $C_{4x}$   \\
      $(\phantom{-}0,\phantom{-}0,-1)$ &       $C_{2a}$   &  $(\phantom{-}1,\phantom{-}0,-1)$  &       $C_{2a}$   &	 $(-1,\phantom{-}1,\phantom{-}1)$  &  $C_{4y}^i$ \\
      $(\phantom{-}0,-1,\phantom{-}0)$ &       $C_{4x}$   &  $(\phantom{-}1,\phantom{-}1,\phantom{-}0)$   &       $C_{4y}$   &	 $(\phantom{-}1,-1,-1)$ &  $C_{2x}$   \\
      $(-1,\phantom{-}0,\phantom{-}0)$ &       $C_{4y}^i$ &  $(\phantom{-}0,\phantom{-}1,-1)$  &       $C_{4x}^i$ &	 $(-1,\phantom{-}1,-1)$ &  $C_{2y}$   \\
                 &                  &  $(\phantom{-}0,-1,\phantom{-}1)$  &       $C_{4x}$   &	 $(-1,-1,\phantom{-}1)$ &  $C_{2z}$   \\
		 & 		    &  $(\phantom{-}0,-1,-1)$ &	   $C_{2f}$   &	 $(-1,-1,-1)$&  $C_{2b}$   \\  
		 & 		    &  $(-1,\phantom{-}0,-1)$ &	   $C_{2b}$   &              &             \\
		 & 		    &  $(-1,\phantom{-}0,\phantom{-}1)$  &	   $C_{4z}$   &		     &             \\
		 & 		    &  $(-1,-1,\phantom{-}0)$ & 	   $C_{2d}$   &		     &             \\
		 &                  &  $(-1,\phantom{-}1,\phantom{-}0)$  &       $C_{4y}^i$ &		     &            

    \end{tabular}
  \end{table}

These lattice correlators are mapped to the reference moving frame by
the corresponding $SU(2)$ $J$-representation of the source and sink
operators. The specific choices of group elements for the mapping are
given in Table~\ref{tab:mf-rots}. The residual moving frames are
related to the reference frames and those listed in
Table~\ref{tab:mf-rots} by momentum inversion. The corresponding
operators and lattice correlators are thus mapped by the parity
transformation $I$.

\subsection{Contractions}
\begin{figure}[htpb]
  \centering
  \includegraphics[width=0.48\linewidth]{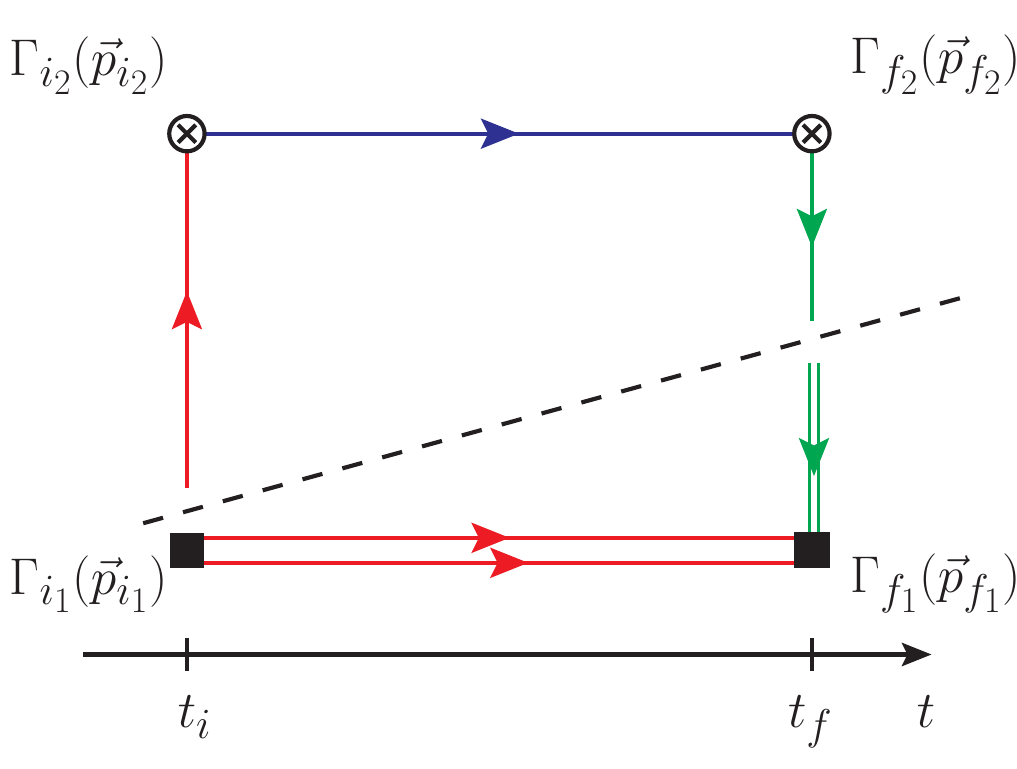}
  \includegraphics[width=0.48\linewidth]{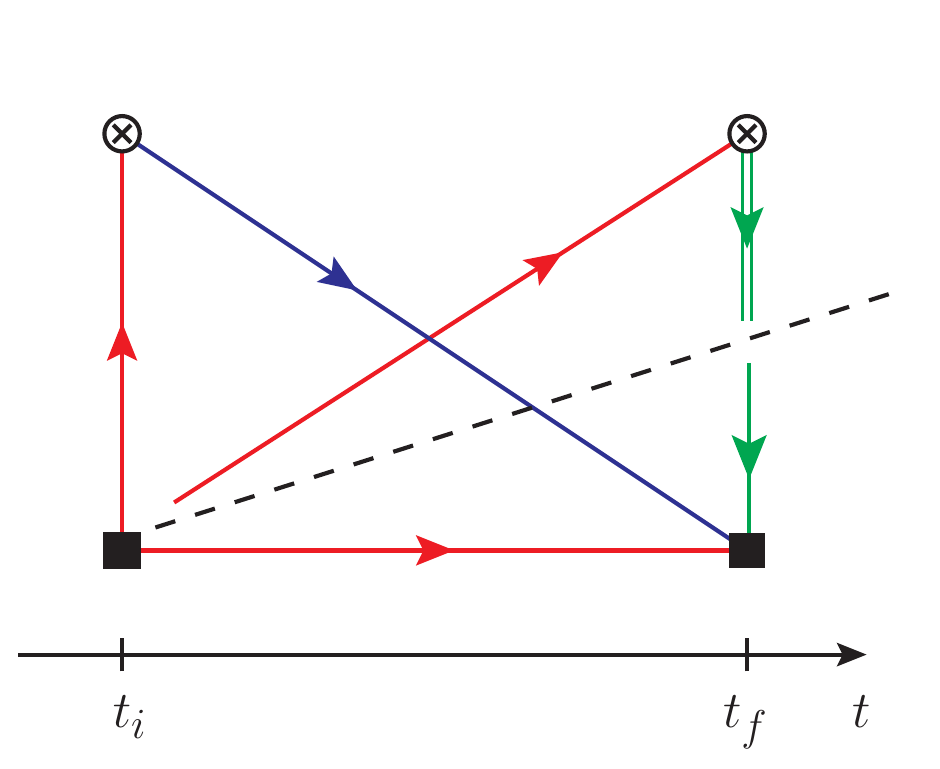}
  \includegraphics[width=0.48\linewidth]{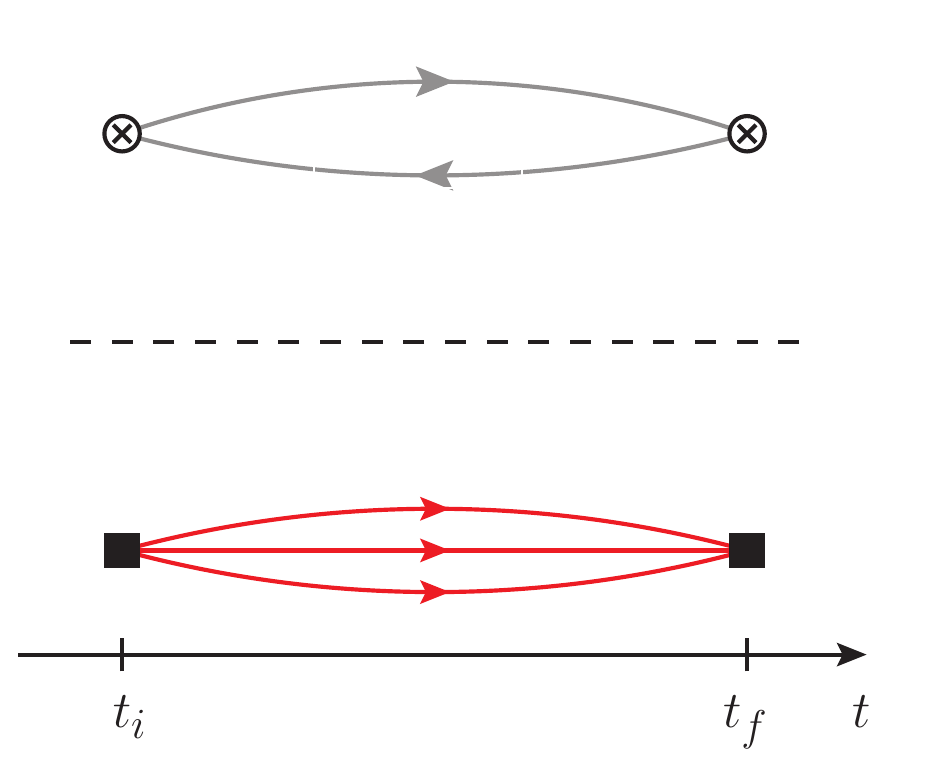}
  \includegraphics[width=0.48\linewidth]{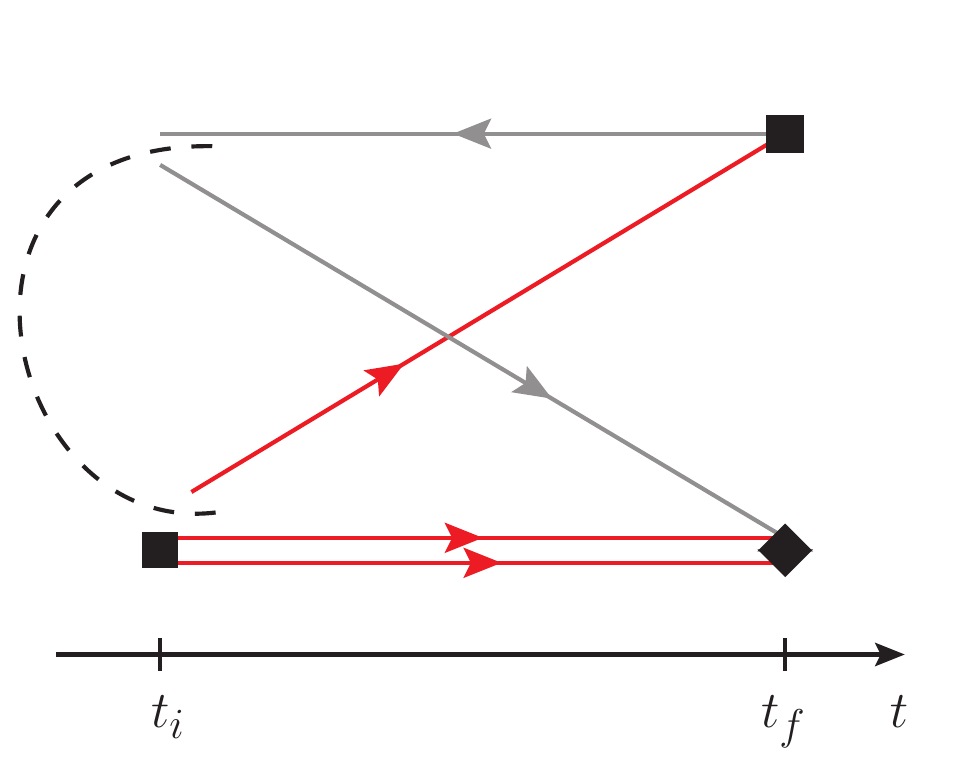}
  \caption{Different types of diagrams showing the contractions of the cost-intensive
    $\pi N$ correlation functions. The dashed line shows the
    applied splitting into diagram factors. As explained in more
    detail in the text, point-to-all propagators are denoted with red
    lines, sequential propagators with blue lines, stochastic sources
    (propagators) with single (double) green lines, and one-end-trick
    stochastic propagator pairs with gray lines.  }
  \label{fig:piN_diags}
\end{figure}

Our method to compute correlation functions of single- and two-particle operators efficiently is
based on the factorization  of the quark flow
diagrams~\cite{Silvi:2021uya}. We show the diagrams that contribute to the $N\pi - N\pi$
two-point function in Fig.~\ref{fig:piN_diags}.

The colors used for the different propagators shown in
Fig.~\ref{fig:piN_diags} denote the method used to evaluate them.
We denote spinor indices by lower case Greek indices,
whereas color indices are lower case Latin indices.

In particular, point-to-all propagators ($S$) denoted with red lines read,
\begin{align}
  S_{\alpha,\beta}^{a,b}(x;x_i) &= { D^{-1 } }_{\alpha,\gamma}^{a,b}(x;x_i)  \nonumber\\ &\includegraphics[width=0.2\textwidth]{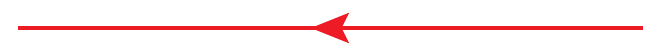}, 
  \label{eq:pta}
\end{align}
where $x_i$, $\beta$, and $b$ are the coordinates, Dirac index, and
color index of the source, respectively. Sequential propagators ($T$) are denoted
with blue lines and are given by,
\begin{align}
  T_{\alpha,\beta}^{a,b}(x; t_{\mathrm{seq}},\pvec_{\mathrm{seq}}; x_i) &=  { D^{-1 }}_{\alpha,\gamma}^{a,c}(x;y)\times \, \nonumber\\ & \left[
    \left(\Gamma_{\mathrm{seq}}\right)_{\gamma\gamma'}\,\epow{i\pvec_{\mathrm{seq}}\yvec} \,\, S_{\gamma',\beta}^{c,b}(t_{\mathrm{seq}},\yvec;\, x_i) 
  \right] \nonumber\\
  &\includegraphics[width=0.2\textwidth]{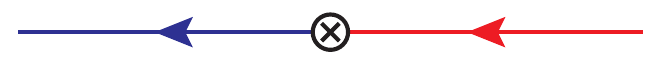},
  \label{eq:seq}
\end{align}
constructed from an inversion of the Dirac operator using a
point-to-all propagator as a source vector (also referred to as the
sequential source) with support on timeslice $t_{\mathrm{seq}}$ and
with a vertex given by Dirac structure $\Gamma_{\mathrm{seq}}$ and
three-momentum insertion $\pvec_{\mathrm{seq}}$.

The stochastic sources ($\eta$) and propagators ($\phi$) are denoted
with green single and double lines respectively,
\begin{align}
  \phi^{r}_{\alpha,a}(x)\,{\eta^{r}_{\beta,b}(y)}^* &= { D^{-1 }}_{\alpha,\gamma}^{a,c}(x;z)\times \nonumber\\
  &\,\left[ \eta^{r}_{\gamma,c}(z) \,{\eta^{r}_{\beta,b}(y)}^*\right]\nonumber\\ 
  &\includegraphics[width=0.2\textwidth]{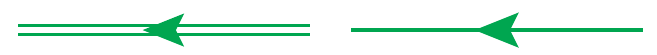},
  \label{eq:stoch}
\end{align}
where $\eta^{r}_{\alpha,a}(x)$ are sources with $Z_2 \times i Z_2$
independent and identically distributed (iid) noise of zero mean and
unit variance,
\begin{align}
  \mathrm{E}\left[ \eta^{r}_{\gamma,c}(z) \right]  &= 0,\nonumber\\
  \mathrm{E}\left[ \eta^{r}_{\gamma,c}(z) \,{\eta^{r}_{\beta,b}(y)}^* \right]
  &= \delta_{\gamma,\beta}\,\delta_{c,b}\,\delta_{\zvec,\yvec}\delta_{t_z,t_y} \,.
  \label{eq:stoch-2}
\end{align}
The expectation value $\mathrm{E[.]}$ is taken over the stochastic
noise index $r$. In practice, we use time-slice noise sources,
i.e. sources with support on a single time-slice,
$\eta^{(r;t_{0})}_{\alpha,a}(x) =
\eta^{r}_{\alpha,a}(x)\delta_{t_x,t_0}$. The gray, one-end-trick
propagators, are constructed using spin-diluted $Z_2 \times iZ_2$
stochastic time-slice sources, appropriately multiplied by a momentum
phase, $\eta^{(r;t_{0};\mu;\vec{p})}_{\alpha,a}(x) =
\eta^{r}_{\alpha,a}(x)\delta_{t_x,t_0}\delta_{\alpha,\mu}e^{i\vec{p}\vec{x}}$
and solution vectors $\phi$ indicated as
$\phi^{(r;t_0;\mu;\vec{p})}_{\beta, b}(y) =
{D^{-1}}^{b,a}_{\beta,\alpha}(y;x)\eta^{(r;t_{0};\mu;\vec{p})}_{\alpha,a}(x)$,

\begin{align}
  \phi^{(r;t_{\mathrm{seq}};\kappa;\pvec_{\mathrm{seq}})}_{\alpha,a}(x)\,\,
  \left( \Gamma_{\mathrm{seq}}\,\gammafive \right)_{\kappa \lambda} \, \, 
  {\phi^{(r;t_{\mathrm{seq}};\lambda;0)}_{\beta',b}(y)}^*\,
  \left(\gammafive\right)_{\beta'\beta} \nonumber\\
   \includegraphics[width=0.2\textwidth]{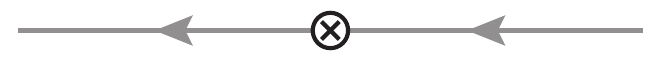}.  &
   \label{eq:oet} 
\end{align}

Analogous to Eq.~\refeq{eq:stoch-2}, the source components are iid
with $Z_2 \times iZ_2$ noise and have zero mean and unit variance.

The diagrams in Fig.~\ref{fig:piN_diags} are further split into
products of building blocks. These cuts are illustrated by the dashed
lines. The quark connected diagrams are split at the source point by
virtue of spin-color dilution, and by the stochastic decomposition of
unity with the stochastic sources and propagators in
Eqs.~\refeq{eq:stoch} and~\refeq{eq:oet}, schematically
$\eta\,\eta^\dagger \approx \mathbb{1}$. The factors are two-
and three-fold propagator products, which are partially reduced in
spin-color space and momentum projected at the pion vertex labeled
$f_2$ in Fig.~\ref{fig:piN_diags} and the nucleon vertex $f_1$.  Each
type of the four diagrams depicted represents several combinations of contractions,
and thus by the factorization multiple contractions benefit from
sharing a small number of common factors.

The Wick contractions for $\Delta - \Delta $ and $\Delta - N\pi$
two-point functions are not found to significantly benefit from this factorization and are therefore calculated 
from point-to-all and point-to-all plus sequential propagators,
respectively.

Both the computation of building blocks and their subsequent
recombination to full correlation functions is carried out on GPUs
using the PLEGMA software package~\cite{plegma}.

\subsection{Quark field smearing}
We apply Gaussian smearing~\cite{Gusken:1989ad} to the source and sink
for all quark propagators. The smearing parameters are
$N_{\mathrm{Gauss}} = 140$ steps with weight $\alpha_{\mathrm{Gauss}}
= 1 $.  The gauge links entering the Gaussian smearing kernel are
APE-smeared \cite{APE:1987ehd} with $N_{\mathrm{APE}} = 50$ steps and
weight $\alpha_{\mathrm{APE}} = 0.4$.  The parameters
$\alpha_{\mathrm{Gauss}}$ and $N_{\mathrm{Gauss}}$ are tuned in order
to approximately give a smearing radius for the nucleon of $
0.5$~fm. 
$\Delta$ ground state energy in the center-of-mass frame, as a
function of the smearing parameters confirmed that these values were
appropriate.

\subsection{Gauge ensemble and statistics}
We use a gauge ensemble generated by the Extended Twisted Mass
Collaboration~\cite{Alexandrou:2018egz} with two degenerate light
quarks with twisted mass parameter tuned to reproduce the pion mass
and the strange and charm quarks ($N_f=2+1+1$) with masses tuned close
to their physical values by matching the kaon meson mass and the ratio
of renormalized quark masses $\left. m_c /
m_s\right|_{\overline{\mathrm{MS}},\mu=2\gev}$, respectively.  The
fermion action is given by twisted mass fermions at maximal twist with
the addition of a Sheihkoleslami-Wohlert ``clover'' term. The gluon
action is the Iwasaki gauge action. The parameters of the ensemble,
denoted as ``cB211.072.64'', are collected in
Table~\ref{tab:ensemble}.
\begin{table}
  \centering
  \caption{Parameters of the ensemble cB211.072.64 used in this work.}
  \label{tab:ensemble}
  \begin{tabular}{cccccc}
    \hline\hline
    $L^3 \times T$   & $a\,[\fermi]$ & $L\,[\fermi]$ & $M_\pi \,[\mev]$ & $M_\pi  L$ & $M_N\,[\mev]$ \\
    \hline
    $64^3 \times 128$ & $0.0801$      & $5.1$         & $ 139.43\,(9)$   & $3.6$      & $ 944\,(10) $ \\\hline
  \end{tabular}
\end{table}

We use $N_{\mathrm{conf}} = 400$ well-separated gauge configurations
and on each configuration generate correlation functions from multiple
source positions ($N_{\mathrm{src}}$) to increase statistics. In
general, we use 64 sources per configuration ($N_{\mathrm{src}}=64$),
except for the specific cases of the nucleon to nucleon ($N$-$N$)
two-point correlation function and the pion-nucleon to pion-nucleon
($\pi N$-$\pi N$) two-point correlation function with both pion and
nucleon at rest, for which we use increased statistics of
$N_{\mathrm{src}}$=256. More details of the statistics, including the
stochastic sources used, are provided in Table~\ref{tab:statistics}.

\begin{table}
  \centering
  \caption{Statistics used in this work. We indicate the number of
    configurations used ($N_{\rm conf}$), and per configuration, the
    number of source points ($N_{\rm src}$) for point-to-all
    propagators, the number of stochastic timeslice sources ($N_{\rm
      stoch}$), and the number of stochastic one-end-trick sources
    ($N_{\rm oet}$).}
  \begin{tabular}{ccccc}
    \hline\hline
    $N_{\mathrm{conf}}$ & $N_{\mathrm{src}}$ & $N_{\mathrm{stoch}}$ & $N_{\mathrm{oet}} $       &   Subset                 \\
    \hline
    $400$               &$256$               & $12$                 & $1$   & \makecell[c]{$N-N$,\\ $\pi N-\pi N$ with $\pvec_\pi=\pvec_N=0$} \\                            
    $400$               & $64$               & $12$                 & $1$   & \makecell[c]{All other correlation functions\\ and kinematic setups} \\\hline                               
  \end{tabular}
  \label{tab:statistics}
\end{table}

\section{Spectral analysis}
\label{sec:spectrum_analysis}
\subsection{Correlator matrices, fits and excited state identification}
We build real symmetric correlation matrices $C_{i j}(t)=\langle
O_{i}(t) \, O_{j}(t_0)^{\dagger}\rangle$, where $O_i(t)$ are the
correlation functions after projecting to the lattice symmetry group,
Eq.~(\ref{eq:projection}).  For extracting the energy levels from the
correlation matrices, we use four methods, which we detail in what
follows.
\subsubsection{Generalized Eigenvalue Problem (GEVP)}
In what we will refer to as the \textit{GEVP method}, we solve the
so-called generalized eigenvalue problem
(GEVP)~\cite{Luscher:1990ck,Blossier:2009kd} for the matrix of
correlation functions
\begin{align}
C_{ij}(t) \, v^{n}_{j}(t_0) &= C_{ij}(t_0) \, v^{n}_j(t_0) \, \lambda^{n}(t,t_0),
\label{eq:gevp}
\end{align}
where $v^n_j(t)$ is the $j^{\rm th}$ component of the $n^{\rm th}$ eigenvector on
time-slice $t$ and $\lambda^{n}(t,t_0)$, the $n^{\rm th}$ eigenvalue of this
GEVP, referred to as the \textit{principal correlator}. In order to
obtain energy eigenvalues, we fit the principal correlator to a single-exponential form
\begin{align}
  f_1(t,t_0) = & A_1e^{-E(t-t_0)}, & t\in[t_{\mathrm{min}}, t_{\mathrm{max}}]. 
\label{eq:gevp-single-exp-fit}
\end{align}

\begin{figure}[htpb]
  \centering
\includegraphics[width=\linewidth]{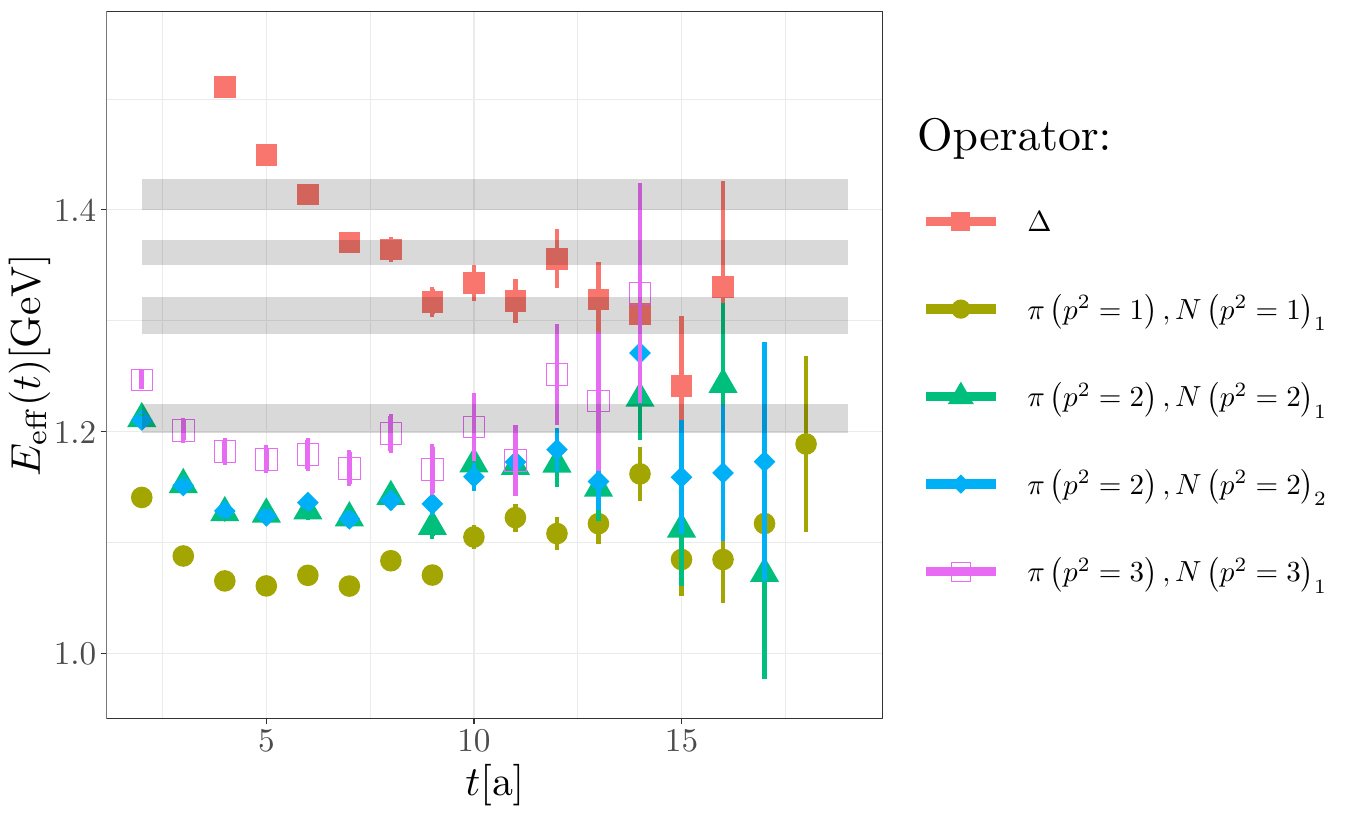}
\caption{The effective mass from each diagonal correlator in the
  center-of-mass frame $H_g$ correlation matrix, using a 5$\times$5
  basis of interpolating operators listed in the legend. The gray
  bands show the energy levels obtained by the GEVP analysis. }
\label{fig:diagonal}
\end{figure}

An example of the analysis using GEVP is shown in
Fig.~\ref{fig:diagonal}, where we plot the effective masses $m^i_{\rm
  eff} = \log \frac{C_{ii}(t)}{C_{ii}(t+1)}$ of the diagonal elements
of the correlation matrix for the case of the center-of-mass frame
$H_g$. The five interpolating operators, that include four momentum
combinations of $\pi N$ and one $\Delta$, are indicated in the legend. 

\subsubsection{Prony Generalized Eigenvalue Method (PGEVM)}
We apply the Prony Generalized Eigenvalue Method
\cite{Fischer:2020bgv,doi:10.1137/0916008}
directly on the principal correlators obtained via
the GEVP method. In PGEVM, we solve the second-level GEVP for the
correlation matrix $C^{(2)}$ defined by $\lambda^{n}$ as
\begin{align}
  C^{(2)}_{ij}(t) &= \begin{pmatrix}
    \lambda^{n}(t  ,t_0)  & \lambda^{n}(t+1,t_0)  \\
    \lambda^{n}(t+1,t_0)  & \lambda^{n}(t+2,t_0)  \\
  \end{pmatrix}_{ij} \,.
  \label{eq:pgem-corelation-matrix}
\end{align}
While with the GEVP method the stability of the ground state from
$\lambda^{n}$ is tested by a conservative choice of
$t_{\mathrm{min}}$, with PGEVM a second, consecutive ground state
projection is applied, which is expected to lead to an earlier onset
of ground state dominance and thus smaller statistical errors.

\subsubsection{Athens Model Independent Analysis Scheme (AMIAS)}
In the AMIAS method~\cite{Alexandrou:2014mka} we perform multi-state
fits directly to the correlation matrix. Namely, the spectral
decomposition of the correlation functions
$C_{ij}(t)=\sum_{n=1}^{n_{\rm max}}A^{(ij)}_n e^{-E_nt} $ is used to construct the
$\chi^2$ function
\begin{align}
\chi^2 &= \sum_{j,k}\sum_{t=t_{\rm min}}^{t_{\rm max}}\frac{\left( C_{jk}(t) -\sum_{n=1}^{n_{\rm max}} A^{(jk)}_n e^{-E_n t}\right)^2}{\sigma_{jk}^2},
\end{align}
where the amplitudes, $A^{(jk)}_n$ and the energies, $E_n$, are fit
parameters and $n_{\rm max}$ is used to truncate the spectral
expansion. The probability distribution function (PDF) for the
complete set of parameters is
$p(A,E)=\frac{1}{Z}e^{-\frac{\chi^2}{2}}$, where the normalization factor  $Z = \int \prod_{n=1}^{n_{\rm max}} d A_n d E_n
e^{-\chi^2/2}$. The estimates for the values of the fit
parameters and their uncertainties are then obtained as the
expectation values and the standard deviations of the corresponding
PDF,
\begin{align}
\bar{E}_k & = \int \prod_{n=1}^{n_{\rm max}} \,  d A_n \, d E_n \,\, E_k \, p(E,A) \,, \\
\sigma^2(E_k) & = \int \prod_{n=1}^{n_{\rm max}} \, d A_n \, d E_n \,\, (E^2_k - \bar{E}^2_k) \, p(E,A)\,.
\label{eq:amias-aveerr}
\end{align}
These integrals are computed using standard Monte Carlo methods. In
AMIAS, one investigates the behavior of the distributions of the fit
parameters of the lower states of interest as the truncation parameter
$n_{\rm max}$ is increased.

\begin{figure}[h]
    \centering
    \includegraphics[width=\linewidth]{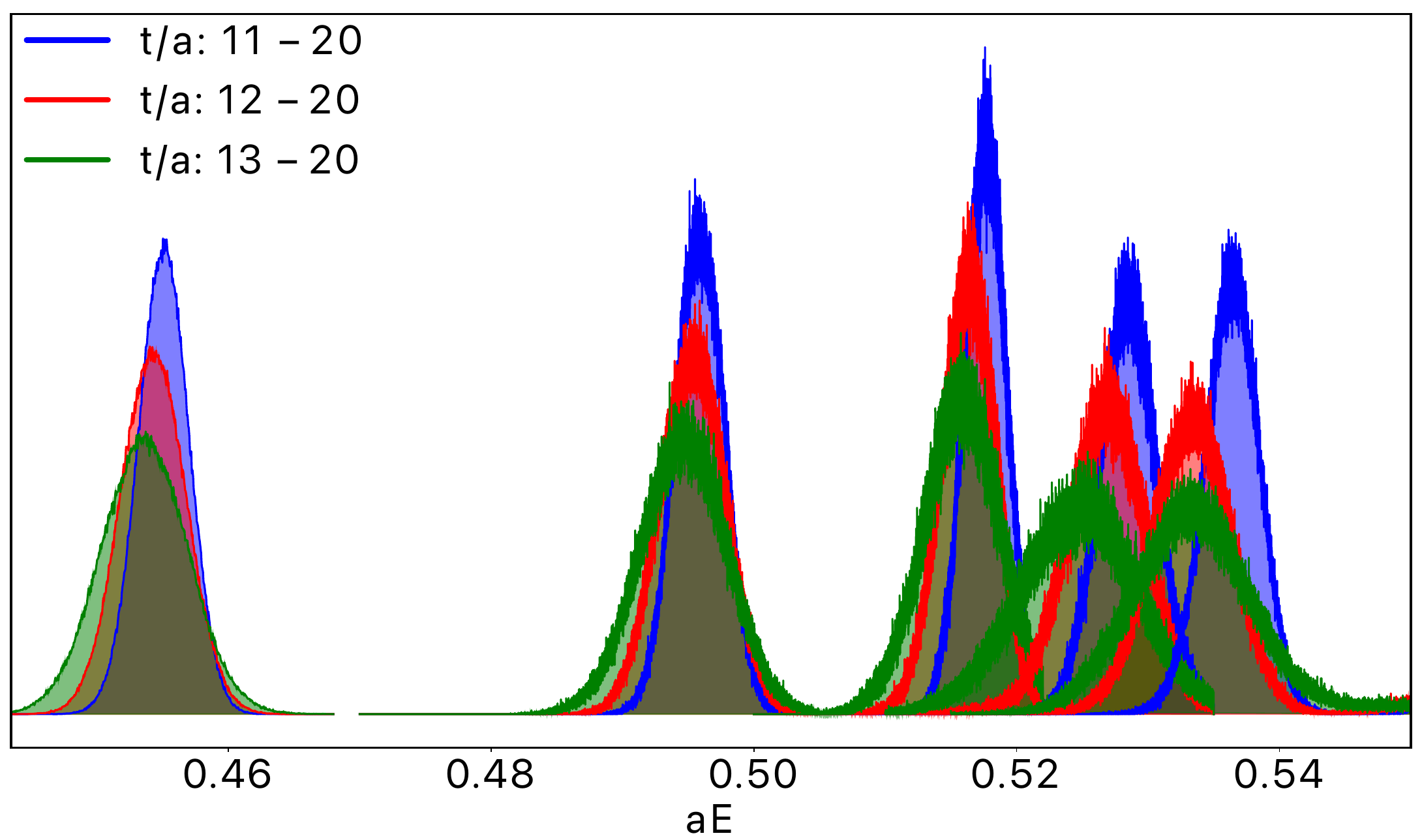}
    \caption{Results on the energy spectrum for the $G_1$ irrep using the AMIAS method. We show the distributions for the energies when $t_{\rm min}/a$=11 (blue curves), 12 (red curves), and 13 (green curves). The correlation matrix is the same as in the $5\times5$ problem used in Fig.~\ref{fig:diagonal}.}
    \label{fig:amias-example}
\end{figure}

As demonstrated in previous applications of this method
\cite{Alexandrou:2014mka,Alexandrou:2017itd,Alexandrou:2019tmk}, at
large values of $n_{\rm max}$, the additional parameters added, to
which $\chi^2$ is insensitive, become irrelevant in the integrals of
Eq.~(\ref{eq:amias-aveerr}) and thus the distributions of the energies
of interest converge, without loss of accuracy.  In this way larger
fit intervals $\left[ t_{\mathrm{min}} \,,\, t_{\mathrm{max}} \right]$
(i.e. smaller $t_{\mathrm{min}}$) can be probed.

An example application of AMIAS is shown in
Fig.~\ref{fig:amias-example}, where the distributions of the energy
levels for the case of the $G_1$ irrep are plotted. As can be seen,
using $n_{\rm max}=5$ all five energy levels are clearly
distinguishable. A similar analysis is carried out for the amplitudes
$A^{(ij)}_n$ and for all irreps considered, to obtain the mean values
and errors of these parameters via Eq.~(\ref{eq:amias-aveerr}). All
results in this work quoted as using the AMIAS method used up to
$n_{\rm max}$=4.

\subsubsection{Ratio Method}
\begin{figure}[htpb]
  \centering
\includegraphics[width=\linewidth]{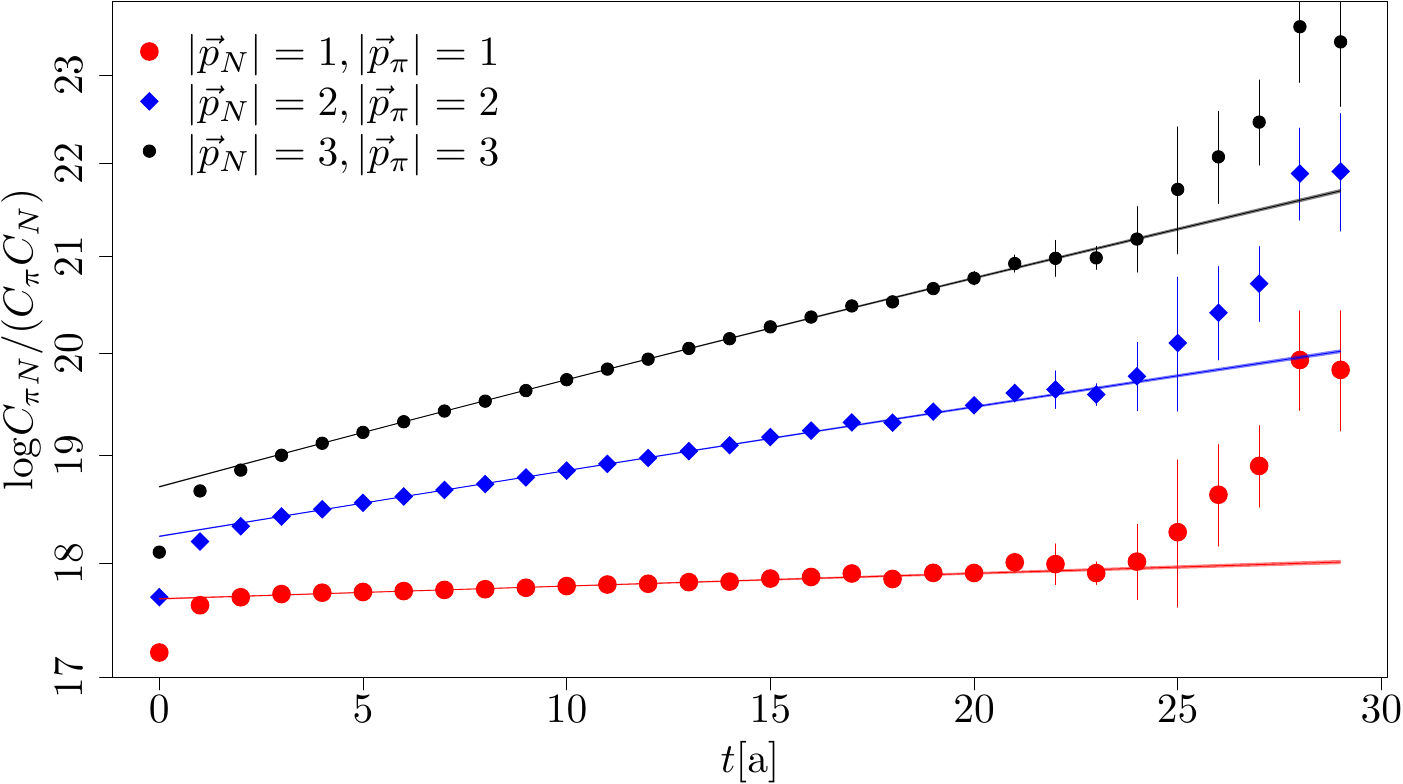}
\caption{The correlation function $C^{R}$ for the ground state in the
  $H_{\mathrm{g}}$ irrep, center-of-mass frame with zero total
  momentum and with each hadron carrying 1 (red circles), 2 (blue
  circles), or 3 (black circles) units of back-to-back momentum. The
  solid lines show linear fits with the values of the energy gap
  $\Delta E_{\vert{\vec{p}_N}\vert=1}=-0.0113(9),\Delta
  E_{\vert{\vec{p}_N}\vert=2}=-0.0612(9),\Delta
  E_{\vert{\vec{p}_N}\vert=3}=-0.103(9)$ and $\chi^2$ per degrees of
  freedom
  $\chi^2_{\vert{\vec{p}_N}\vert=1}=0.53,\chi^2_{\vert{\vec{p}_N}\vert=2}=0.52,\chi^2_{\vert{\vec{p}_N}\vert=3}=0.4$,
  respectively}
\label{fig:ratio_Hg}
\end{figure}

Following Ref. \cite{Bulava:2016mks} we take 
the principal correlator from the GEVP analysis, and fit the
energy shift with respect to a given non-interacting energy level. In
practice this is done by taking the ratio of the principal correlator
$\lambda^n(t,t_0)$ 
from Eq. \refeq{eq:gevp}
to the product of single hadron correlation functions with the appropriate
momenta, given by
\begin{align}
C^{R}_{\vec{p}_{\mathrm{N}},\vec{p}_{\mathrm{\pi}}}(t) &= 
\frac{
  \lambda^{\vec P, \Lambda,n}(t,t_0)
}{
  C_N(t,\vec{p}_{\mathrm{N}}) \cdot C_\pi(t,\vec{p}_{\mathrm{\pi}})
}\,,
\label{eq:ratio} \\
\vec{P} &= \vec{p}_{\mathrm{N}}+\vec{p}_{\mathrm{\pi}} \,.
\nonumber
\end{align}
We will refer to this approach as the \textit{Ratio method}. 
The leading energy dependence of the ratio in Eq. \refeq{eq:ratio}
follows as
\begin{align}
  C^{R}_{\vec{p}_{\mathrm{N}},\vec{p}_{\mathrm{\pi}}}(t) &\overset{\mathrm{large~} t}{\propto}
  \frac{
    \epow{-E^{\vec P,\Lambda,n}\,t}
  }{
    \epow{-E^{\pvec_{\mathrm{N}}, \Lambda_N,0}\,t}
    \epow{-E^{\pvec_{\pi},\Lambda_\pi,0}\,t}
  }
  \nonumber \\
  &= 
   \epow{-\left( E^{\vec P,\Lambda,n} - E^{\pvec_{\mathrm{N}}, \Lambda_N,0} - E^{\pvec_{\mathrm{N}}, \Lambda_\pi,0} \right) \,t}
  \nonumber \\
  &= \epow{-\Delta E^{\vec P,\Lambda,n}(\vec{p}_{\mathrm{N}},\, \vec{p}_{\mathrm{\pi}})\,t},
  \label{eq:ratio-2}
\end{align}
where $E^{\pvec_{\mathrm{N}}, \Lambda_N,0}$ and
$E^{\pvec_{\pi},\Lambda_\pi,0}$ are the ground state energy of the
nucleon and pion two-point function in the relevant irreps,
respectively. Taking the ratio reduces substantially the excited state
contribution in a given principal correlator, and especially for small
energy shifts $\Delta E^{\vec P,\Lambda,n}(\vec{p}_{\mathrm{N}},\,
\vec{p}_{\mathrm{\pi}})$ allows to determine the latter with
significantly increased statistical precision. For large $t$ the
logarithm of $C^R(t)$ will converge to a linear function, with the
slope corresponding to the energy shift $\Delta E^{\vec
  P,\Lambda,n}(\vec{p}_{\mathrm{N}},\, \vec{p}_{\mathrm{\pi}})$
relative to the non-interacting $\pi\,N$ energy.
\begin{align}
 \log\left(   C^{R}_{\vec{p}_{\mathrm{N}},\vec{p}_{\mathrm{\pi}}}(t) \right)  
 &\overset{\mathrm{large~} t}{\propto}
 \mathrm{const} - \Delta E^{\vec P,\Lambda,n}(\vec{p}_{\mathrm{N}},\, \vec{p}_{\mathrm{\pi}}) \cdot t \,.
  \label{eq:ratio-3}
\end{align}
An example of such an analysis
is shown in Fig.~\ref{fig:ratio_Hg}, for the case of the $H_g$ irrep,
which corresponds to the center-of-mass frame with total momentum zero (see
Table~\ref{tab:irreps}). The ratio is applied to the $\pi\,N$ case for
three different relative momenta $\vec{p}_{\mathrm{N}},\, \vec{p}_{\mathrm{\pi}}$ 
between the nucleon and pion. 
From
the energy shift obtained by the slope, we reconstruct the interacting
two-hadron energy level
by shifting back the energies using the
continuum dispersion relation
\begin{align}
  E^{\vec P,\Lambda,n} &= \Delta E^{\vec P,\Lambda,n}(\vec{p}_{\mathrm{N}},\, \vec{p}_{\mathrm{\pi}})  
  \nonumber \\
  &\qquad + \sqrt{m_N^2 +  \vec{p}_{\mathrm{N}}^2 } + \sqrt{m_\pi^2 +  \vec{p}_{\pi}^2 }\,.
  \label{eq:ratio-4}
\end{align}
Eq. \refeq{eq:ratio-4} has the added advantage, that high-precision estimates for the 
nucleon and pion mass can be employed.
 
Using different single-hadron momenta for pion and nucleon with same total momentum
to determine the same energy shift and interacting energy level is 
part of our systematic error analysis.

\subsubsection{Correlation matrix basis selection}
For our analysis, we progressively add interpolating fields to the
correlation matrix used, selecting the most appropriate basis by
checking the stability of the spectrum. In particular, the steps we
follow are:
\begin{enumerate}
  \item We include all the relevant single-nucleon and pion momentum
    combinations and occurrences and perform a GEVP analysis.
  \item Based on the eigenvectors obtained from the full GEVP
    analysis, we restrict the GEVP to using the interpolating
    operators that dominate the first few energy levels.
  \item Starting from the smaller GEVP of the previous step, we
    gradually extend the basis. The interpolating operators to be
    added are chosen by observing the effective mass of their diagonal
    correlators, and whether they yield higher energy states than
    already seen in the smaller GEVP. As we add these interpolating
    operators, we check that the statistical errors of the lower-lying
    energy spectrum do not deteriorate.
\end{enumerate}

\begin{figure}[htpb]
  \centering
  \includegraphics[width=\linewidth]{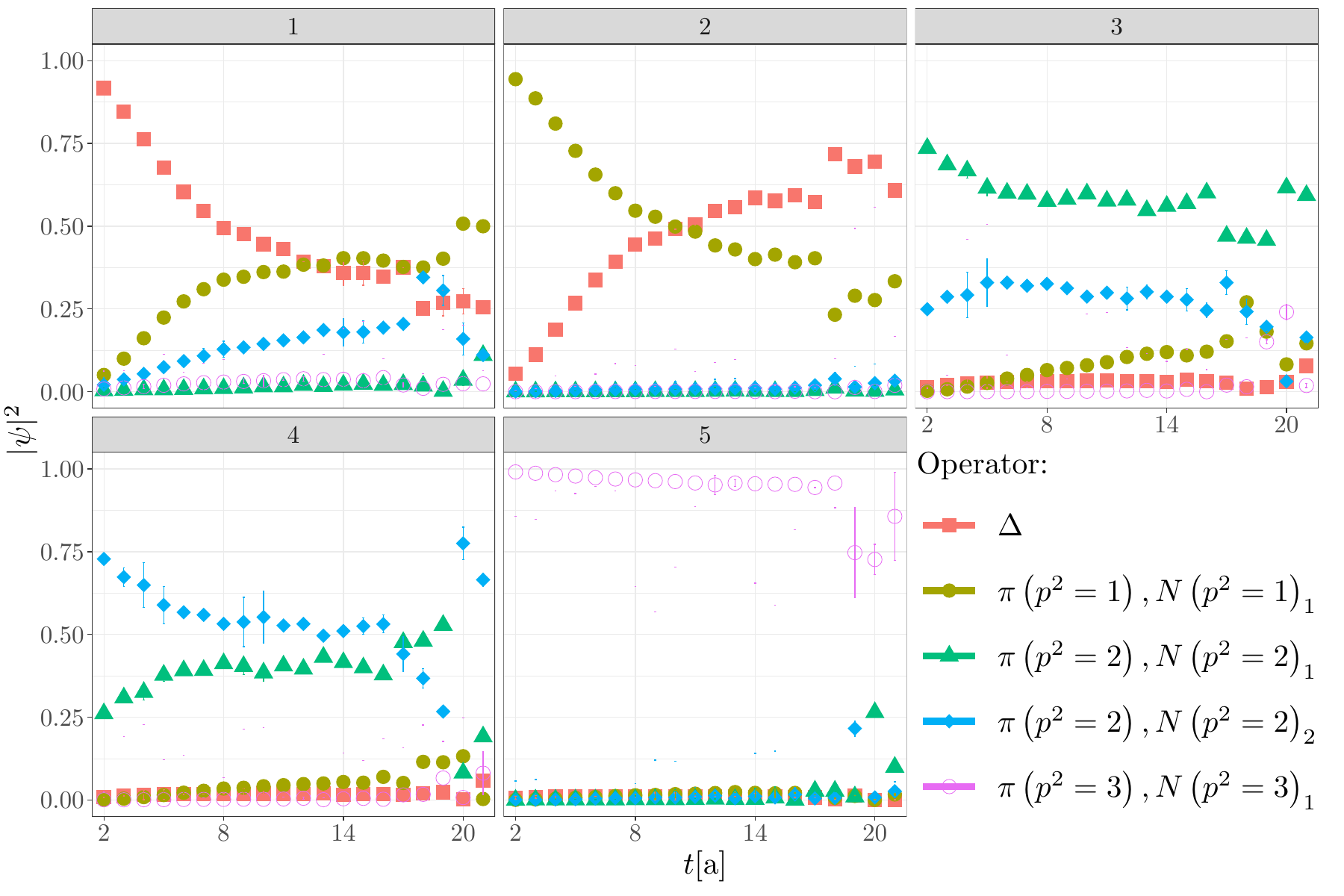}
  \caption{The components of the first five eigenvectors obtained from
    a GEVP in the centre-of-mass frame irrep $H_g$. The number in the
    title of each figure corresponds to $n$ in Eq.~\ref{eq:overlap},
    while the different symbols denote the interpolating operator
    ($O_\alpha$ in Eq.~\ref{eq:overlap}) as indicated in the legend.}
\label{fig:decomposition}
\end{figure}

In Fig.~\ref{fig:decomposition}, we illustrate the basis selection
process using an example taken from the GEVP method and the
center-of-mass frame irrep $H_g$. What is plotted are the five
components of each eigenvectors obtained via the GEVP, which is solved
on each time-slice. These correspond to the overlaps of each
interpolating field used with the lowest-lying energy state.  For a
certain choice of eigenvector $v^{n}$ obtained via the GEVP, and
interpolating operator number $\alpha$, the overlap is defined as
\begin{align}
  |\psi|^2 &= |v^{n}_{\alpha}|^2 = |\brackets{n\,\vert\,O^\dagger_{\alpha}\,\vert\,0}|^2.
  \label{eq:overlap}
\end{align}
The eigenvectors are normalized to unity,
$\sum\limits_{\alpha}\,|v^{n}_{\alpha}|^2 = 1$ for all $n$. From
Fig.~\ref{fig:decomposition}, we see that the first two eigenvectors of
the GEVP are largely dominated by the amplitudes from two
interpolators, $O_{\Delta}^{H_g}$ and $O_{N\pi}^{H_g}(\pvec,-\pvec)$
with $\pvec^2 = 1$.

We note that this same basis was used for Fig.~\ref{fig:diagonal}
discussed earlier, where we see that the pion-nucleon interpolating
field decreases the ground state energy in the particular channel we
consider here. We use this basis to extract energy levels for the 
analysis that follows.

\subsection{Lattice spectrum and stability test}
\begin{figure}[h]
\center
  \includegraphics[width=0.9\linewidth]{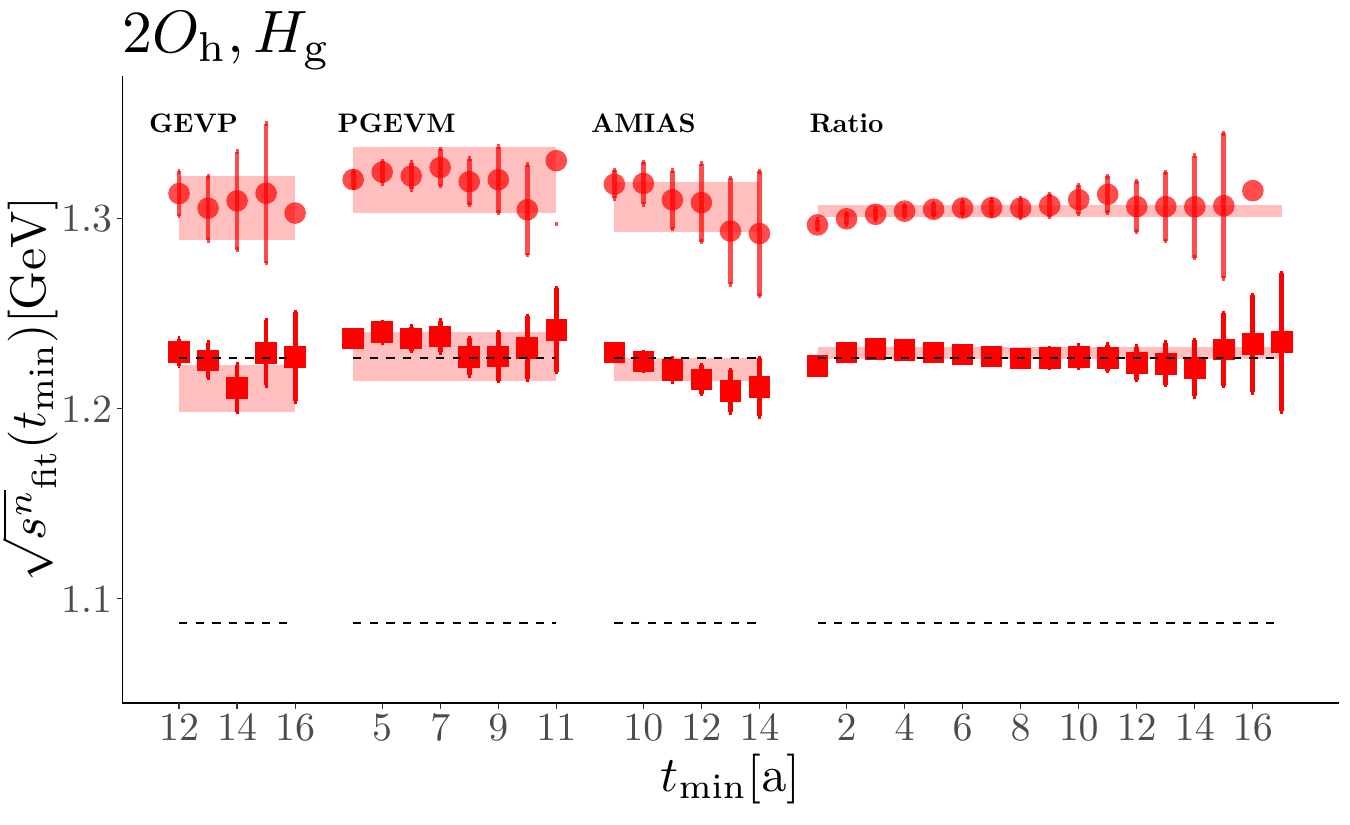}
  \caption{Energy levels for the case of the $H_g$ irrep. The set of
    points, from left to right, correspond to results from the GEVP
    method, the PGEVM method, AMIAS, and the ratio method. Dashed
    lines indicate non-interacting levels for comparison. The red
    bands correspond to values listed in Table~\ref{tab:energy_values}
    for GEVP, PGEVM, and AMIAS, and to the values with the smallest
    $t_\mathrm{min}$ listed in Table~\ref{tab:energy_ratio} for the
    ratio method.}
  \label{fig:stability-example}
\end{figure}

The four methods detailed above are complementary in the way the
excited state contamination is treated. Thus, by comparing the results
obtained among them, we can check the robustness of our observed
energy levels. The comparison is carried out for all irreps
considered, observing the stability of the results as we increase the
initial fit-range ($t_{\rm min}$). An example is shown in
Fig.~\ref{fig:stability-example}, where the four methods are compared
for the specific case of the $H_g$ irrep. As can be seen in this plot,
the PGEVM and ratio methods provide stable results at smaller $t_{\rm
  min}$ compared to the standard GEVP. Furthermore, the statistical
errors carried by the ratio method confirm our expectation that the
correlations between numerator and denominator in Eq.~(\ref{eq:ratio})
help in reducing the statistical fluctuations in the energy shift from
the non-interacting energy. This same analysis is carried out for all
irreps, with the corresponding plots given in
appendix~\ref{appendix:stability-plots}. By observing the stability of
the fitted masses as $t_{\rm min}$ is increased, as well as the
$\chi^2/{\rm d.o.f}$ of the fit, we indicate our selected values for
each level and for each method with the bands and in
appendix~\ref{appendix:tables-results} collect the results for GEVP,
PGEVM, and AMIAS in Table~\ref{tab:energy_values} and for the ratio
method in Table~\ref{tab:energy_ratio}, where we also include results
for two larger $t_{\rm min}$ values that we use in a model averaging
for our final result in Sec.~\ref{sec:ratio}.

\begin{widetext}

\begin{figure}[h]
  \centering
  \includegraphics[width=1\textwidth]{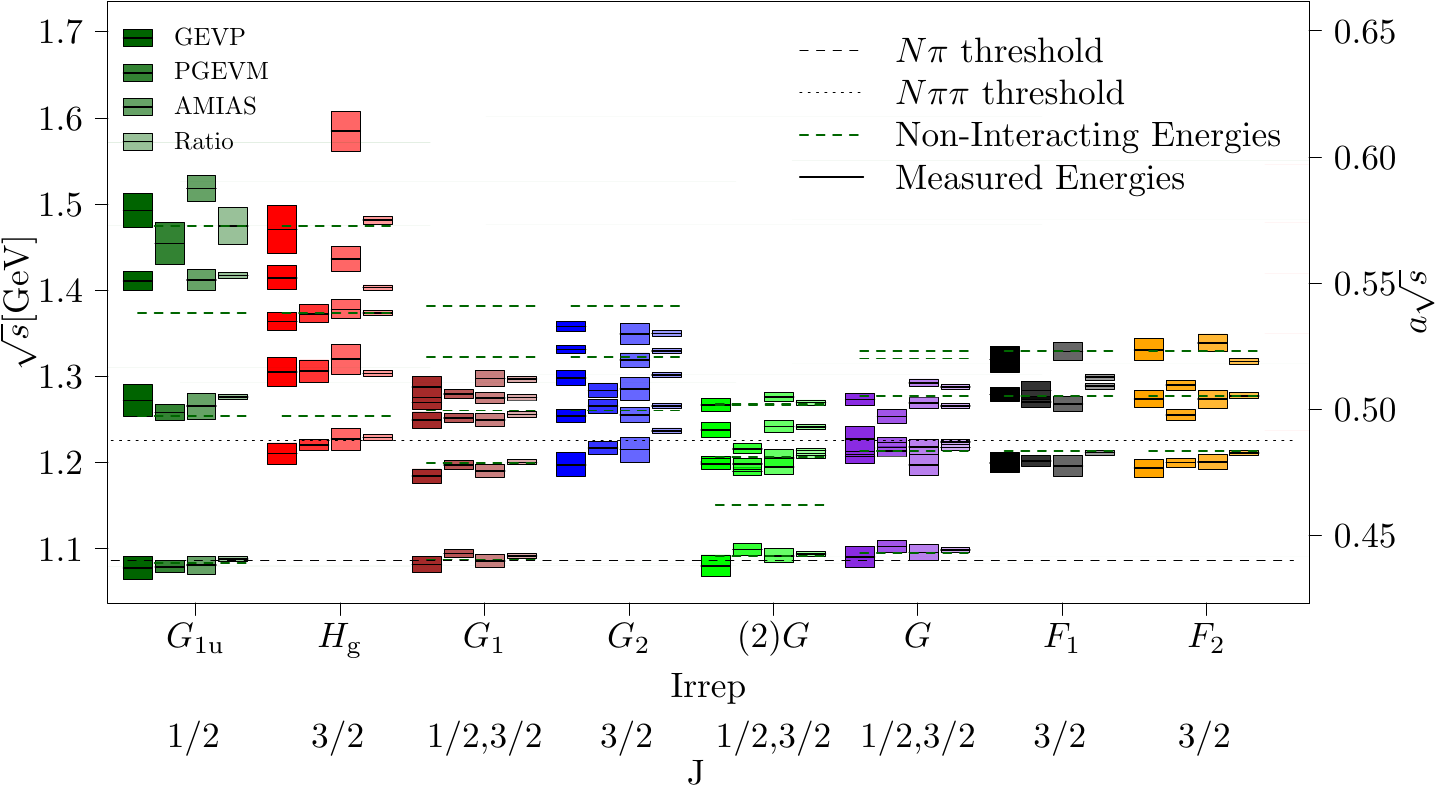}
  \caption{The $\pi N$ interacting two-hadron energy levels obtained
    by our analysis. For each irrep and total angular momentum $J$,
    indicated on the x-axis, we include results using the four methods
    employed, namely, from left to right, the GEVP method, the PGEVM
    method, AMIAS, and the ratio method, with the band height
    indicating our estimated uncertainties as explained in the
    text. On the left y-axis we indicate the energy levels in physical
    units, while on the right y-axis in lattice units. The gray dashed
    and dotted lines spanning the entire x-axis correspond to the $\pi
    N$ and $\pi\pi N$ thresholds, namely 1080 and 1220~MeV,
    respectively. The green, thicker dashed lines correspond to the
    non-interacting $\pi N$ energy levels permitted for each irrep and
    for the volume used in this work.}
  \label{fig:spectrum_survey}
\end{figure}

\end{widetext}

In Fig.~\ref{fig:spectrum_survey}, we collect all the energy levels
extracted from all irreps considered, using all four methods. The $\pi
N$ threshold ($E_{2-\rm thr}$) and $\pi\pi N$ ($E_{3-\rm thr}$) are
shown for comparison at 1080~MeV and 1220~MeV
respectively. Furthermore, for each irrep we indicate the permitted
non-interacting energy levels that correspond to the lattice volume of
the ensemble used. For GEVP and PGEVM, the band, indicating the
uncertainty of the energy levels, contains both statistical and a
systematic error from the fit range variation. For AMIAS the
systematic error from varying the fit range is negligible. For the
ratio method we include only statistical errors here. A dedicated
discussion of how we obtain the systematic errors of the ratio method
is given in Sec.~\ref{sec:ratio}.

As can be seen, the GEVP, PGEVM, and AMIAS methods yield comparable
statistical errors, while the ratio method yields smaller statistical
errors, as expected given the previous discussion of
Fig.~\ref{fig:stability-example}. In general, we can identify several
energy levels that are incompatible with non-interacting energy
levels, however the statistical errors carried by our results,
combined with the proximity of the $\pi \pi N$ threshold and the first
non-interacting energy, make clearly identifying the $\Delta$
resonance rather challenging, even with the large statistics of
$\mathcal{O}$(10$^4$) used in this work. Our analysis, when compared
to the analogous mesonic system of the $\rho$
resonance~\cite{Fischer:2020yvw}, highlights the increased
requirements for extracting resonance parameters of systems that
include baryons and when using physical point ensembles.

All data plotted in Fig.~\ref{fig:spectrum_survey} are included in
Tables~\ref{tab:energy_values} and~\ref{tab:energy_ratio} of
appendix~\ref{appendix:tables-results}.

\section{Scattering parameters}
\label{sec:scattering_analysis}

The scattering amplitude near the resonance is well described by a
Breit-Wigner type resonance.  We thus parameterize the
$\Delta$-resonance phase shift, which via the L\"uscher quantization
condition then predicts the finite volume energy spectrum.  The
lattice spectrum we have determined is then fitted to the prediction,
and from the minimization of $\chi^2$ we extract the optimal
parameters of the resonance given our spectrum results. In practice,
we determine the roots of the quantization condition of
Eq.~(\ref{eq:lqc}) for the given set of resonance parameters and
construct a correlated, non-linear $\chi^2$, given by

\begin{align}
  \chi^2= & \sum_{i\in\left({\Gamma},n\right)}\sum_{j\in\left({\Gamma},n\right)} \,\, \xi_i \,  w_{ij}^{-1} \, \xi_j\nonumber\\
  \xi_i = & aE^{i,lat}_{\mathrm{cms}} - aE^{i}_{\mathrm{cms}}\left(L,M_R,\Gamma_R,a_0\right),
\end{align}
where $w_{ij}$ is the covariance matrix between the lattice data for
$aE^{i,lat}_{\mathrm{cms}}$ and $aE^{j,lat}_{\mathrm{cms}}$ estimated
using jackknife resampling. To determine the errors of the fit
parameters via the jackknife procedure, we perform the minimization of
$\chi^2$ in each jackknife sample. In
Table~\ref{tab:qc_fit_systematics_test}, we collect the results
obtained for the scattering parameters when using the energy levels
determined via the GEVP, PGEVM, or AMIAS methods. For the ratio
method, which we will use to quote our final values, we carry out a
more thorough analysis of the errors that we describe later in this
manuscript.  For the results in
Table~\ref{tab:qc_fit_systematics_test}, we either restrict to using
only $P$-wave dominated irreps, with partial wave $J=3/2$ and
$\ell$=1, which involves including five levels in the $\chi^2$
minimization, or we use $S$- and $P$-wave dominated irreps, with two
partial waves $(J,\,\ell)$=(3/2, 1) or (1/2, 0), thus including in
total 14 energy levels. For the latter case, we estimate the
scattering length from the combined $S$- and $P$-wave fits via
Eqs.~(\ref{eq:delta-s-wave}) and ~(\ref{eq:delta-p-wave}). In this
work we restrict to providing the scattering length only for this
channel, leaving the isospin 1/2 case for a future publication.

  \begin{table}[h]
    \caption{ Results for the scattering parameters, namely the
      resonance mass, $M_R$, resonance width, $\Gamma_R$, and
      scattering length, $M_\pi a_0$, using the L\"uscher quantization
      condition and energy levels determined via the GEVP, PGEVM and
      AMIAS methods. First three rows when using $P$-wave only and
      last three rows when using $S-P$ wave dominant irreps.
      \label{tab:qc_fit_systematics_test}}
    \begin{center}
      \begin{tabular}{ccccc}\hline\hline
        \multirow{2}{*}{Method} & \multicolumn{2}{c}{Breit-Wigner parameters} & \multirow{2}{*}{$M_\pi a_0$} & \multirow{2}{*}{$\chi^2/$dof} \\
                                &   $M_R$ [MeV] & $\Gamma_R$ [MeV]             &                              &                               \\\hline
                                \multicolumn{5}{c}{Fit to ($J,\ell$) = (3/2, 1); No. of $\sqrt{s}$ points = 5}\\
     GEVP  &  1249(42) &  180(240)& -         & 0.61  \\
     PGEVM &  1274(63) &  160(260)& -         & 0.55  \\
     AMIAS &  1271(40) &  160(260)& -         & 0.47  \\
     \\\hline
                                \multicolumn{5}{c}{Fit to ($J,\ell$) = (1/2, 0) and (3/2, 1); No. of $\sqrt{s}$ points = 14}\\
    GEVP   &  1262(36) &  190(135)& -0.20(7)  & 0.82  \\
    PGEVM  &  1270(36) &  142(207)& -0.04(24) & 0.29  \\
    AMIAS  &  1274(11) &  180(180)& -0.20(14) & 1.28  \\
      \end{tabular}
      \end{center}
  \end{table}

As can be seen from Table~\ref{tab:qc_fit_systematics_test}, results
when using the three methods are overall compatible and within
statistical errors are consistent with the experimental determinations
of the $\Delta$ mass and width.  However, the statistical errors for
the resonance width are large and do not permit a significant
comparison with experiment. We, therefore, opt to using the ratio
method, presented below, to quote our final results for the resonance
parameters. We note that for the results in
Table~\ref{tab:qc_fit_systematics_test}, an interpolation method was
used to accelerate the minimization of $\chi^2$ in each jackknife bin,
described in detail in appendix~\ref{appendix:optimization}.

\subsubsection{Results using the ratio method}
\label{sec:ratio}
The most accurately determined energy levels are obtained using the
ratio method and are employed to obtain our final values of the
resonance parameters. Given these smaller errors, a thorough analysis
of the sources of systematic errors is merited, and we, therefore,
consider the following in our fits when using the ratio method:
\begin{enumerate}
  \item We consider two different ranges of center-of-mass energies,
    namely (a) including only energy levels below $N\pi\pi$ threshold
    ($E_{3-\mathrm{thr}}$), which leads to including 12 energy levels,
    and (b) energy levels up to $\Delta \, \pi$
    ($E^{\prime}_{3−\mathrm{thr}})$, which leads to including 14
    levels. The latter is defined by the onset of the rise of the
    inelasticity in the $J=3/2,\,\ell = 1$ pion nucleon scattering
    channel \cite{Ronchen:2012eg}.
  \item We explore the partial wave dependence, i.e. we consider fits
    with energy levels coming only from $P$-wave dominant irreps,
    which leads to including 5 levels, and energy levels having also
    an $S$-wave contribution, which leads to the combinations
    mentioned in the previous item. We attempted to include higher
    partial waves but this led to prohibitively large statistical
    errors on the parameters.
  \item The $S$-wave contribution is entirely parameterized by the
    scattering length, which can be obtained directly from the ratio
    of correlators used in the ratio method, as presented in more
    detail in Sec.~\ref{sec:scattering length} below. We either
    perform fits using this direct determination of the scattering
    length or leaving the scattering length free as an additional fit
    parameter.
  \item We use three different fit ranges for the energy levels,
    i.e. three values of $t_\mathrm{min}$, that indicatively span
    between $t_\mathrm{min}\simeq$~0.3~fm and 1~fm. The smallest
    $t_\mathrm{min}$ is determined from the onset of the plateau in
    the ratio method and is that corresponding to the band in
    Fig.~\ref{fig:stability-example} and
    Figs.~\ref{fig:stability-G1u}-\ref{fig:stability-F2}. The largest
    $t_\mathrm{min}$ is determined from the onset of the plateau in
    the single exponential fit to the principal correlator from the
    GEVP. An intermediate $t_\mathrm{min}$ is also used between these
    two values.
  \item We vary the non-interacting energy level, i.e. that of the
    denominator in Eqs.~\refeq{eq:ratio}, \refeq{eq:ratio-2} in the
    ratio method. We found in our fits, that the energy levels most
    sensitive to this variation are the ones for the $H_g$ irrep. We,
    therefore, perform a separate analysis for the $H_{g}$ energy
    levels using the ratio method with pion-nucleon states with three
    values of back-to-back momenta, namely $\pvec_{\mathrm{N}}^2 =
    \pvec_{\pi}^2 = \left( \frac{2\pi}{L} \right)^2\cdot \left\{
    1,\,2,\,3 \right\}$, as in Fig.~\ref{fig:ratio_Hg}.
\end{enumerate}

Considering these variations, our analysis yields 45 results, over
which we quantify our systematic uncertainty. The results are
tabulated in Table~\ref{tab:qc_fit_systematics} of
appendix~\ref{appendix:systematics}.

\begin{figure}[h]
  \centering
  \includegraphics[width=1\linewidth]{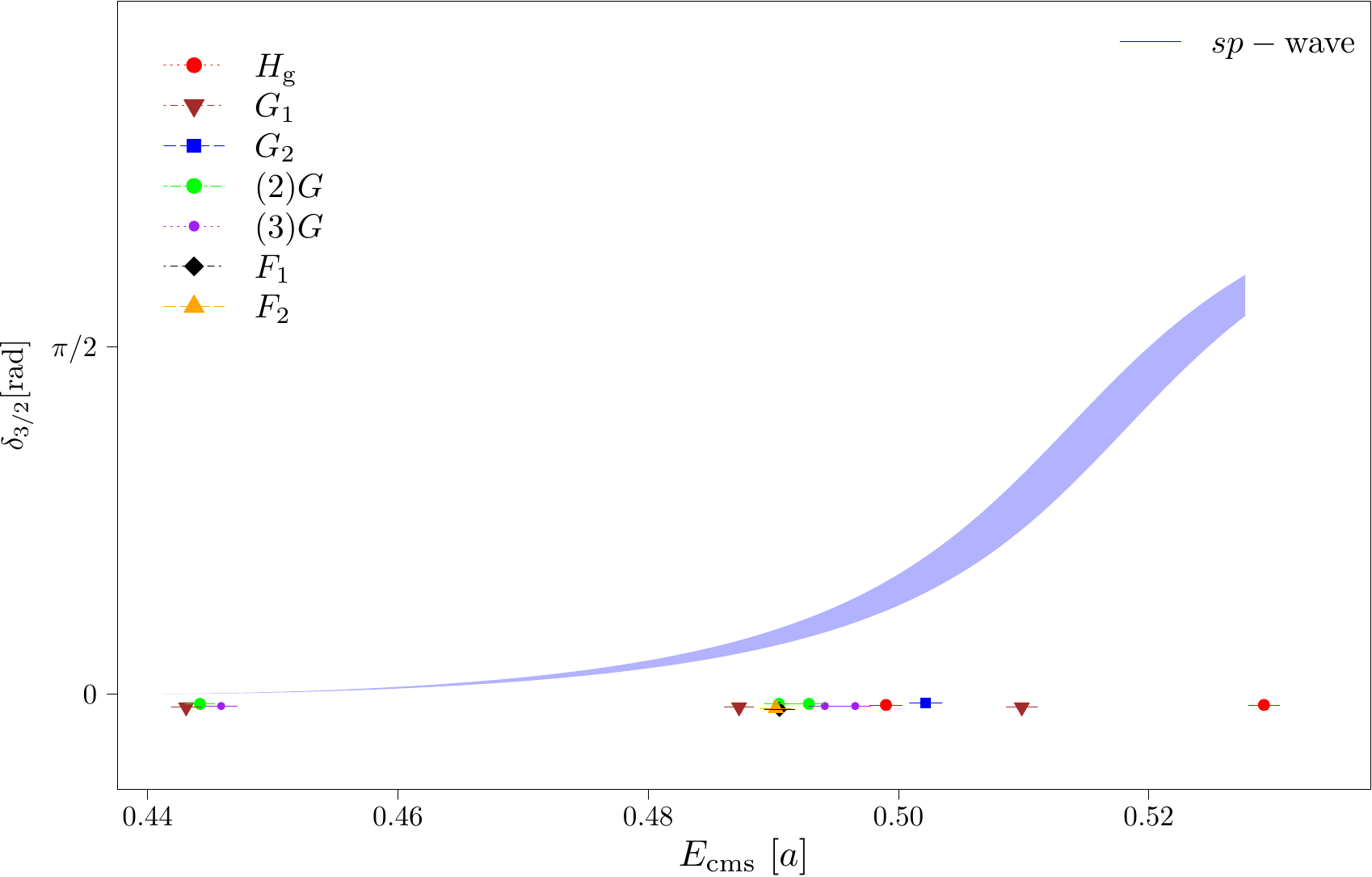}
  \caption{The $P$-wave phase-shift as a function of the invariant
    mass $E_{\rm cms} = \sqrt{s}$. The error band is determined using
    jackknife resampling. The points with horizontal errorbars show
    each fitted energy level included its jackknife errorbar.}
  \label{fig:phase-1}
\end{figure}

The results of the different fits are model averaged according to
Ref.~\cite{Neil:2022joj} to derive a combined statistical and
systematic error.  As an example, in Fig.~\ref{fig:phase-1}, we show
our result for the $P$-wave phase-shift for one of the 45 fits that
has a significant contribution to the model average.

\begin{figure}[!h]
    \centering
    \includegraphics[width=0.9\linewidth]{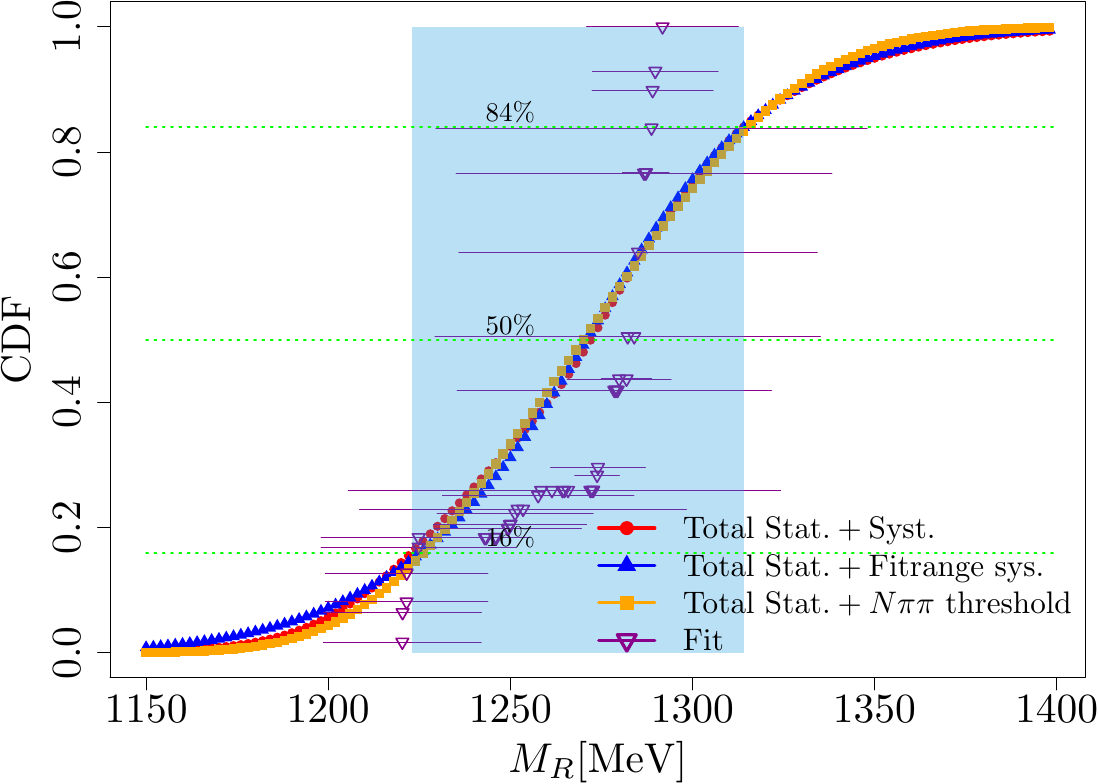}\\[3ex]
    
    \includegraphics[width=0.9\linewidth]{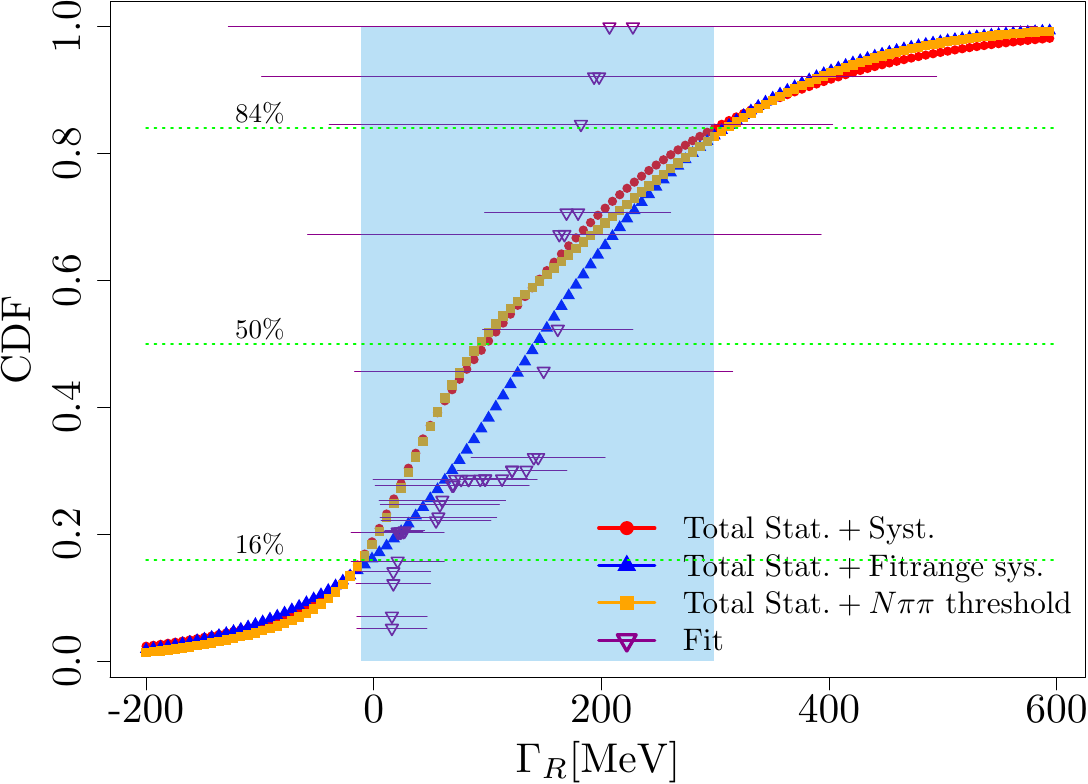}
    \caption{The cumulative probability distribution function for the
      model averaging carried out to obtain the resonance mass $M_R$
      (top) and the resonance width $\Gamma_R$ (bottom). We
      distinguish between systematics arising from the energy-level
      fit range choice (blue triangles), 
      considering energy levels also slightly above $N\pi\pi$
      threshold (brown squares) and including all five sources of
      systematic errors as explained in the text (red small
      circles). For the resonance mass all three curves fall on top of
      each other, while for the width the yellow squares are on top of
      the red circles.  Each dark magenta point corresponds to the
      central value and statistical error of one particular fit with
      $y$-axis corresponding to the numerically determined CDF. The
      vertical band shows the range of $M_R$ (top) and $\Gamma_R$
      (bottom) between 16\% and 84\% which we take as our total
      error.}
    \label{fig:cdf}
  \end{figure}

The cumulative distribution function (CDF) for the resonance parameters
obtained from the 45 different fits is shown in
Fig.~\ref{fig:cdf}. Our final result for the
resonance mass and width obtained through model averaging are 
\begin{center}
$M_R = 1269\,(39)_{\rm{Stat.}}(45)_{\rm{Total}}$~MeV \\
$\Gamma_R = 144\,(169)_{\rm{Stat.}}(181)_{\rm{Total}}$~MeV,
\end{center}
respectively, where the first error is the statistical error and the
second is the total, combining statistical and systematic errors.  For
the individual contributions of the different sources of systematic
error, we refer to Table~\ref{tab:sys_estimating}. The systematic
error for each source of uncertainty ($a$) is estimated via 
\begin{equation}
\sigma^a_{\rm{sys}}=\sqrt{\sum_{i=1}^{N_a}{\mathcal O_i}^2p_i-\left(\sum_{i=1}^{N_a}{\mathcal O}_ip_i\right)^2},
\end{equation}
where ${\mathcal O}_i$ and $p_i$ are averaged over all other systematics other than $a$ and $N_a$ is the number of variations for the given systematic error.
\begin{table}[h]
\caption{Individual contributions to the total systematic uncertainty,
  obtained as explained in the text. Enumeration of the source of
  systematic uncertainty ($a$) follows that of Sec.~\ref{sec:ratio}.
  \label{tab:sys_estimating}}
\centering
\begin{tabular}{cccc}\hline\hline
  \multirow{2}{*}{Criterion ($a$)} & \multirow{2}{*}{$N_a$} & $\Delta M_{R}$ & $\Delta \Gamma_R$\\
  & & [MeV] & [MeV] \\\hline
Inclusion of energies above $N\pi\pi$ (1) & 2 & 35 & 63 \\
$P$- or $P$- and $S$-wave (2) & 2 & 1 & 1  \\
Fit or fixed scattering length (3) & 2 & 26 & 10 \\
Fit range (4) & 3 & 2 & 7 \\
Momenta in $H_g$ (5) & 3 & 4 & 6 \\
\end{tabular}
\end{table}

For the $I=3/2$ $S$-wave scattering length, we
obtain from the combined $S$- and $P$-wave fit of the lattice spectrum
\begin{center}
$M_\pi a_0 = -0.16\,(11)$.
\end{center}

For further illustration, in Fig.~\ref{fig:cdf}, we show 
separately the systematics that originate from varying the fit ranges, from using different
non-interacting levels in the ratio method and considering the two center-of-mass energy ranges. The fact that these curves
collapse onto the curve that corresponds to the total statistical plus
systematic error indicates that our dominant source of error is
statistical.

  \subsubsection{Direct extraction of the scattering length}
  \label{sec:scattering length}
\begin{figure}[h]
  \centering
\includegraphics[width=0.495\textwidth]{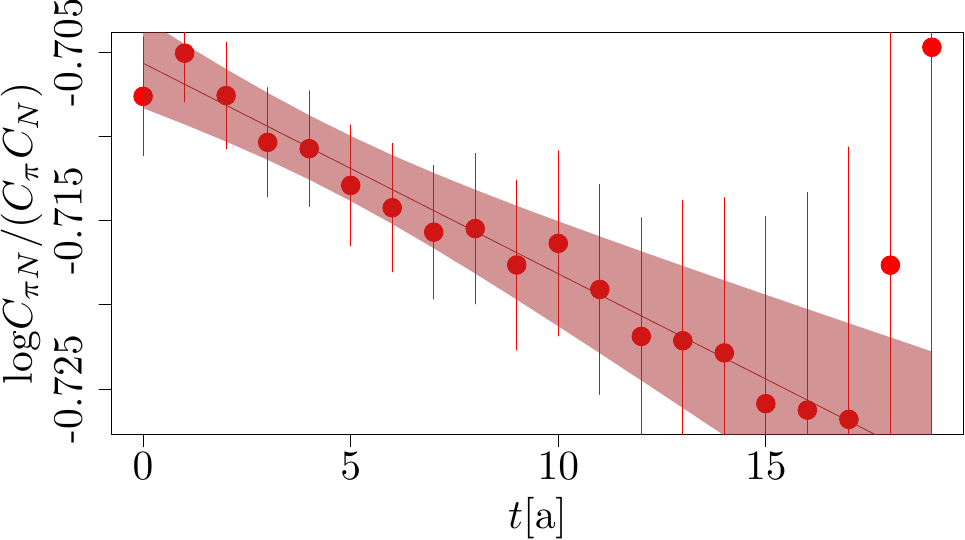}
\caption{The ratio method as applied to the first level of the
  $G_{1\mathrm{u}}$ irrep. With red band we indicate our fit in the
  range $\frac{t}{a}\in\left[3,17\right]$.}
\label{fig:scatt_ratio}
\end{figure}

The scattering amplitude in the $S$-wave around the $\pi\,N$ threshold
can be well described by the leading order effective range expansion
using a single parameter, the scattering length. As an alternative to
the analysis of the previous section, by which the scattering length
is extracted from the phase shift, we can extract this quantity from
the energy shift $\Delta E$ obtained directly from the ratio
method. In particular, correlation functions computed with increased
statistics, indicated in the first row of Table~\ref{tab:statistics},
allow for a high-statistics calculation of the levels in the
$G_{1\mathrm{u}}$ irrep.  Using the first level of this irrep and the
effective range expansion, we determine the scattering length
from~\cite{Luscher:1986pf}
\begin{align}
  \Delta E\cdot L=-\frac{2\pi}{\mu_{\pi N}L}\frac{a_0}{L}\left(1+c_1\frac{a_0}{L}+c_2\left(\frac{a_0}{L}\right)^2\right)+{\mathcal O}(L^{-5}).
  \label{eq:effective range expansion}
\end{align}
This extraction of $\Delta E$ from the ratio method benefits from the
smaller statistical errors associated with the energy shifts. The data
obtained for the ratio and the resulting fit are shown in
Fig.~\ref{fig:scatt_ratio}.

For the scattering length we obtain
\begin{align}
M_\pi a_0 = -0.13\,(4).
\end{align}
A comparison of our result to other work in the literature is provided
in Table~\ref{tab:scatt_length}. In particular, we compare our results
to a recent lattice calculation using $N_f=2+1$ clover improved
fermions at the heavier-than-physical pion mass of 200~MeV.  We also
compare to three other determinations, namely from pionic
atoms~\cite{Hoferichter:2015hva,Hoferichter:2015tha,Hoferichter:2015dsa,Ditsche:2012fv} using the values of the scattering
lengths updated in Ref.~\cite{Hoferichter:2023ptl}, from unitarized
chiral perturbation theory~\cite{Bruns:2010sv}, and from a global fit
to low-energy pion-nucleon cross-section
data~\cite{RuizdeElvira:2017stg}. Within our quoted statistical
uncertainty, our result is consistent with the latter three results.

\begin{table}[h]
  \caption{Scattering length in the isospin $3/2$ pion-nucleon
    channel. We compare our result, obtained via
    Eq.~(\ref{eq:effective range expansion}), to a lattice calculation
    using 200~MeV pion mass~\cite{Bulava:2022vpq}, a calculation using
    pionic atom data~\cite{Hoferichter:2023ptl}, a calculation using
    unitarized ChPT~\cite{Bruns:2010sv}, and a phenomenological
    determination via global fits to pion-nucleon cross-section
    data~\cite{RuizdeElvira:2017stg}. The error in our determination
    is only statistical.}
  \label{tab:scatt_length}
\centering
\begin{tabular}{ccr@{.}l}\hline\hline
                                           & $M_{\pi}(\mathrm{MeV})$ & \multicolumn{2}{c}{$M_\pi a_0^{3/2}$} \\\hline
 This work                                 & 139                     & -0 & 13(4)                            \\
 Bulava et. al~\cite{Bulava:2022vpq}       & 200                     & -0 & 2735(81)                         \\
 Pionic atoms~\cite{Hoferichter:2023ptl}   & 140                     & -0 & 0865(18)                         \\
 Unitarized ChPT~\cite{Bruns:2010sv}       & 140                     & -0 & 0894(17)                         \\
 Phenomenology~\cite{RuizdeElvira:2017stg} & 140                     & -0 & 0867(35)                         \\
 \hline
\end{tabular}
\end{table}

\section{Conclusions}
\label{sec:conclusions}
Using an ensemble of twisted mass fermions simulated with two
degenerate light quarks, and strange and charm quarks with masses
tuned to their physical values, we determine the Breit-Wigner
resonance parameters of the lowest lying resonance in the $\pi\,N$
$I(J^P) = \frac{3}{2}(\frac{3}{2}^+)$ channel. To our knowledge, this
is the first such lattice study using physical point simulations. We
use a large number of measurements in order to tackle the expected
increase of the statistical uncertainties in the meson-baryon
correlation functions.  To determine the energy levels, we form
correlation matrices of one and two-particle correlation functions
after an appropriate group-theoretic projection to the relevant
lattice irreducible representations. We employ four methods to extract
the energy levels from the correlation functions, including the
standard GEVP method and variants thereof.  While the four methods
yield consistent results, the ratio method, where we determine the
energy gap between interacting and non-interacting energy levels via
appropriate ratios of two-hadron to single-hadron correlation
functions, yields considerably smaller statistical errors.

Restricting our analysis to energy levels obtained via the ratio
method, we solve the L\"uscher quantization condition to extract the
resonance mass and width, varying our fits to probe systematic
uncertainties. For the scattering length, we use a direct approach,
forming a ratio between correlation functions with increased
statistics to obtain the energy shift between non-interacting and
interacting states explicitly. For the resonance parameters we find
the values,
\begin{align}
    M_\pi a_0 &= -0.13\,(4),\nonumber\\
    M_R &= 1269\,(45)\,\mathrm{MeV}  \mathrm{~and}\nonumber\\
    \Gamma_R &= 144\,(181)\,\mathrm{MeV},
\end{align}
for the scattering length in the $3/2$ isospin channel, the resonance
mass, and the resonance width, respectively. Our result for the
scattering length compares well with phenomenological
determinations~\cite{Schopper:1983hnv} and determinations from
ChPT~\cite{Hoferichter:2023ptl,RuizdeElvira:2017stg,Bar:2019igf,Bruns:2010sv}.
A recent lattice study using simulations with 200~MeV
pions~\cite{Bulava:2022vpq} yields a value between 1.4 and 2.7 times
larger than our value when taking into account our statistical
uncertainties, a factor which is compatible with the ratio of pion
masses used. Our result for the resonance mass is compatible with the
expected 1230 - 1234~MeV values quoted by the
PDG~\cite{Workman:2022ynf}, while our value for the resonance width
has large uncertainties, requiring more lattice input to be determined
with significance.  As a first application of the L\"uscher method to
the $\pi N - \Delta$ channel at the physical point, this calculation
was restricted to a single ensemble, which does not allow for a
complete assessment of lattice systematic errors, such as cut-off
effects. The current study, and in particular the level of precision
obtained given the statistics used, paves the way for future
calculations with multiple physical point ensembles allowing for
controlled continuum and infinite volume extrapolations.

\begin{acknowledgments}
FP acknowledges financial support by the Cyprus Research and
Innovation foundation under contracts with numbers
EXCELLENCE/0918/0129 and EXCELLENCE/0421/0195. CA acknowledges support
by the Cyprus Research and Innovation foundation under contract with
number EXCELLENCE/0421/0043.  MP acknowledges support by the
Sino-German collaborative research center CRC 110. This work was
supported by a grant from the Swiss National Supercomputing Centre
(CSCS) under projects with IDs s702, s954, and s1174. We also
acknowledge PRACE for awarding us access to Piz Daint, hosted in
Switzerland. The authors gratefully acknowledge the Gauss Centre for
Supercomputing e.V. (www.gauss-centre.eu) for funding this project by
providing computing time through the John von Neumann Institute for
Computing (NIC) on the GCS Supercomputer
JUWELS-Booster~\cite{DBLP:journals/corr/abs-2108-11976} at J\"ulich
Supercomputing Centre (JSC).
\end{acknowledgments}

\bibliography{bibliography}

\newpage

\appendix

\section{Plots comparing GEVP, PGEVM, AMIAS, and the ratio method}
\label{appendix:stability-plots}

\begin{figure}[!h]
  \centering
  \includegraphics[width=0.8\linewidth]{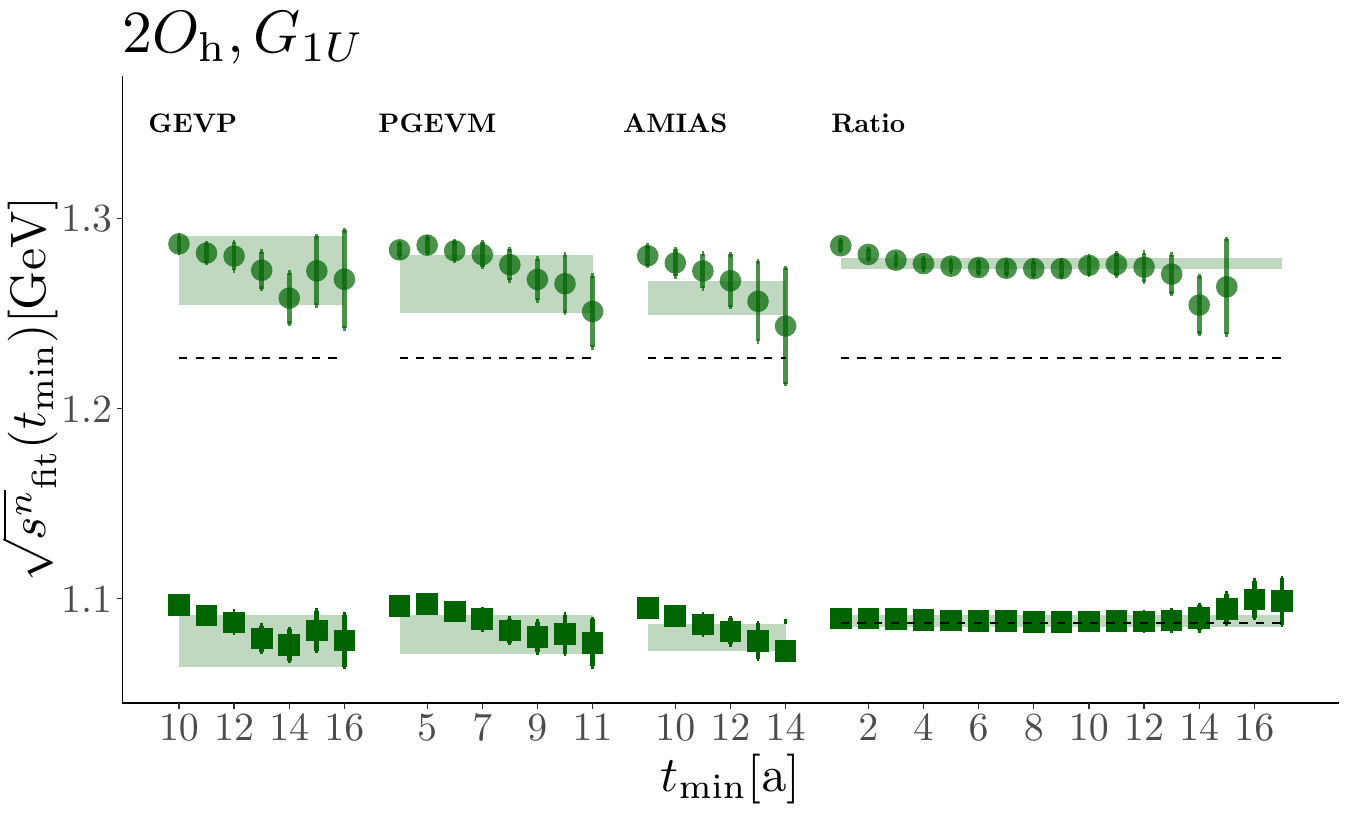}
  \caption{$N\pi$ energy level fits using the GEVP, PGEVM, AMIAS, and
    Ratio methods for the irrep $G_{1u}$.  The colored bands
    correspond to our final selection that is used in
    Fig.~\ref{fig:spectrum_survey}.}
  \label{fig:stability-G1u}
\end{figure}

\begin{figure}[!h]
  \centering
  \includegraphics[width=0.9\linewidth]{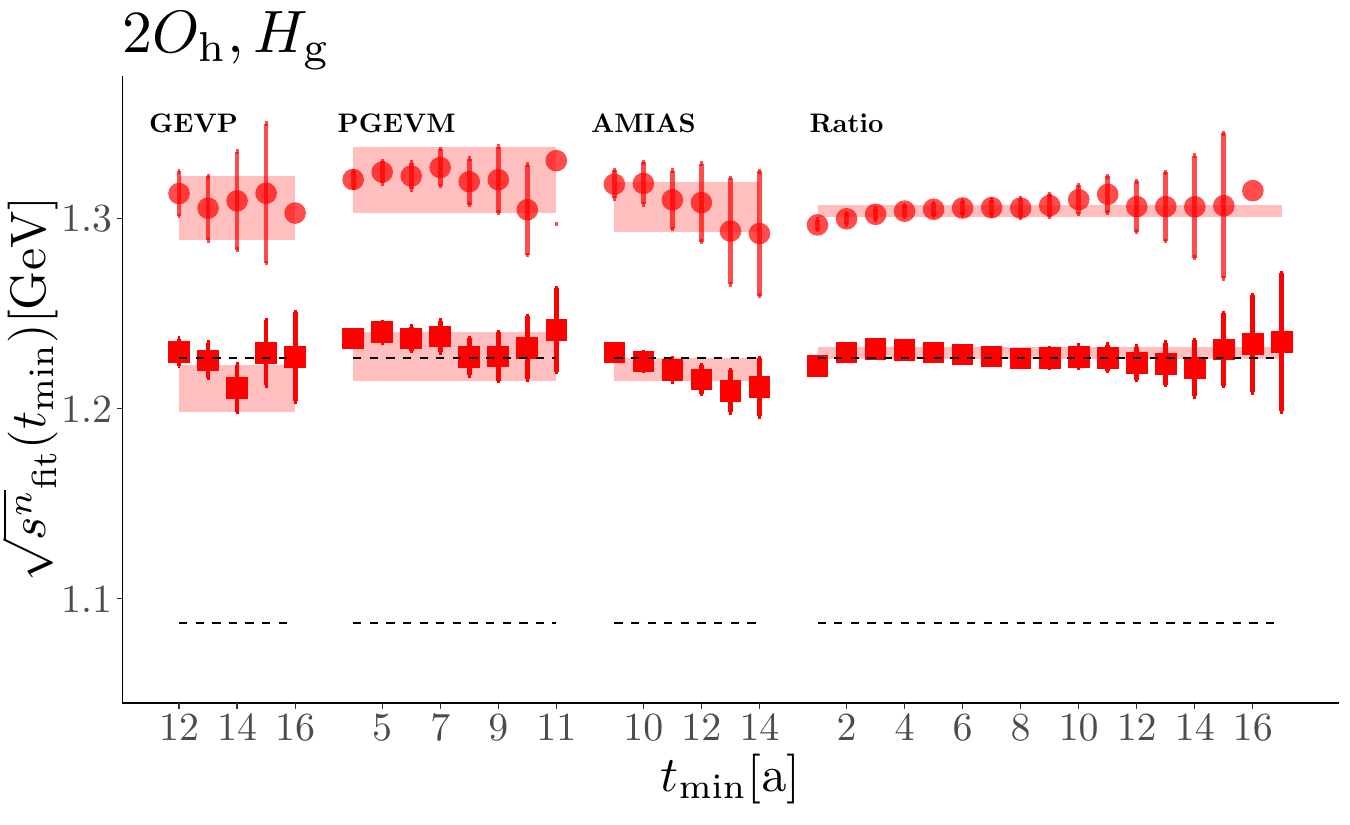}
  \caption{Same as in Fig.~\ref{fig:stability-G1u} but for irrep $H_g$.}
  \label{fig:stability-Hg}
\end{figure}

\begin{figure}[!h]
  \centering
  \includegraphics[width=0.9\linewidth]{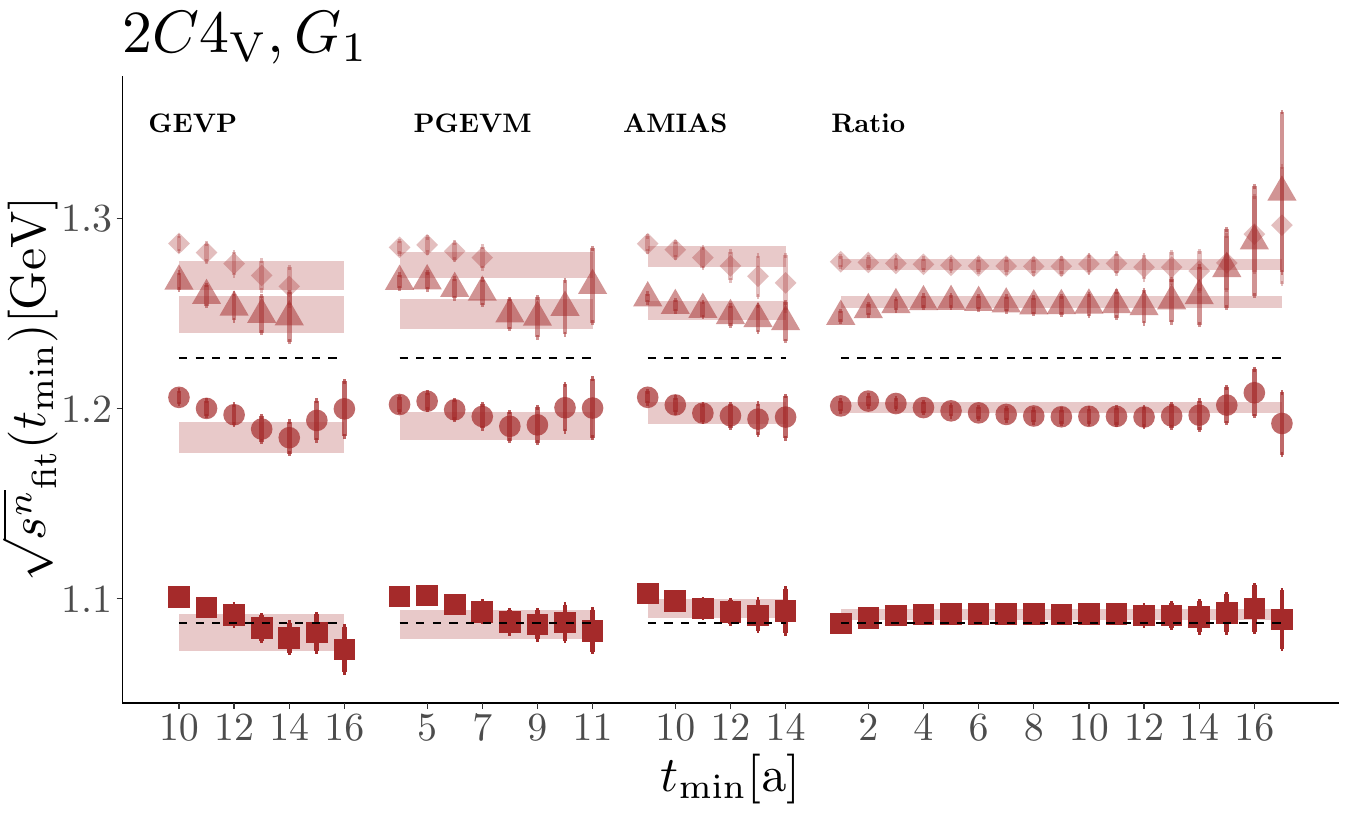}
  \caption{Same as in Fig.~\ref{fig:stability-G1u} but for irrep $G_1$.}
  \label{fig:stability-G1}
\end{figure}

\begin{figure}[!h]
  \centering
  \includegraphics[width=0.9\linewidth]{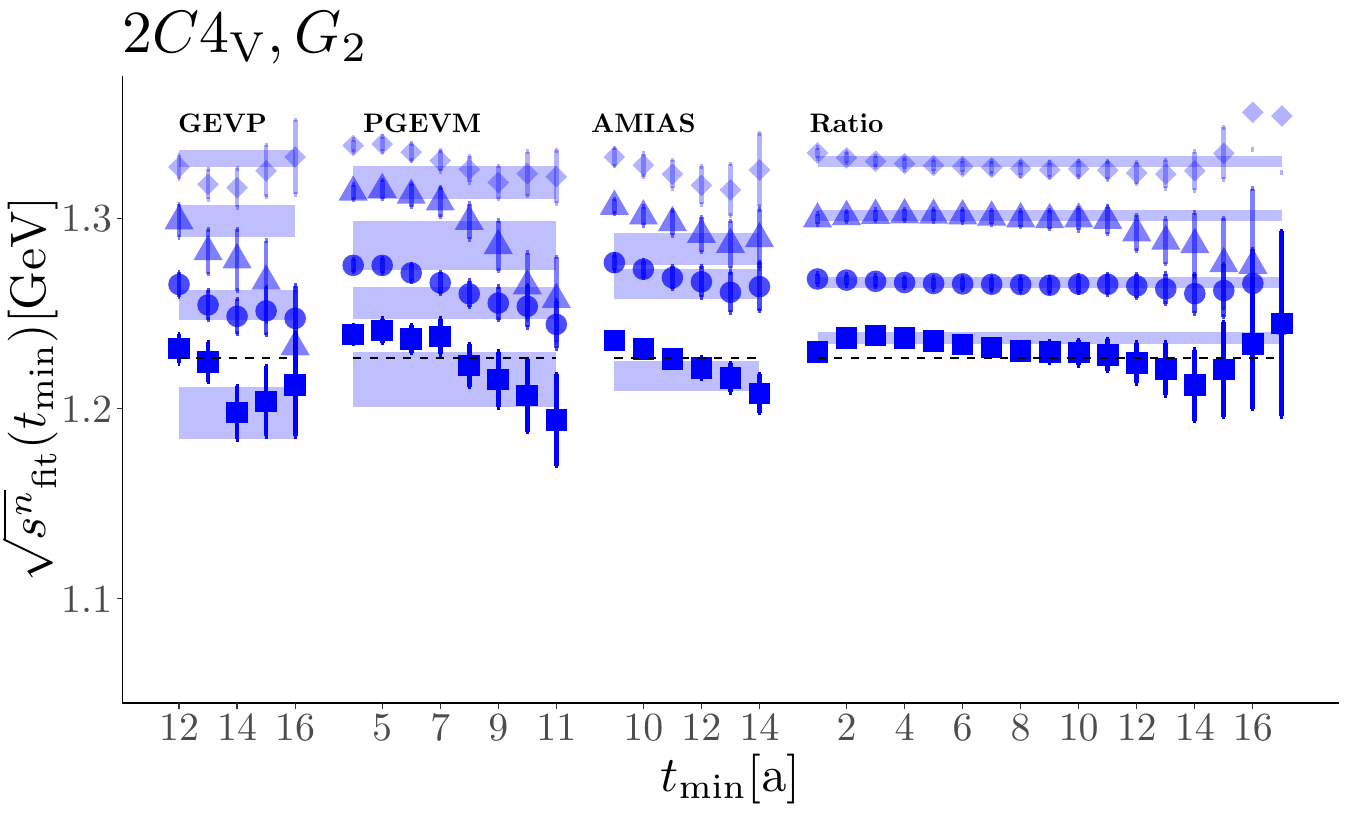}
  \caption{Same as in Fig.~\ref{fig:stability-G1u} but for irrep $G_2$.}
  \label{fig:stability-G2}
\end{figure}

\begin{figure}[!h]
  \centering
  \includegraphics[width=0.9\linewidth]{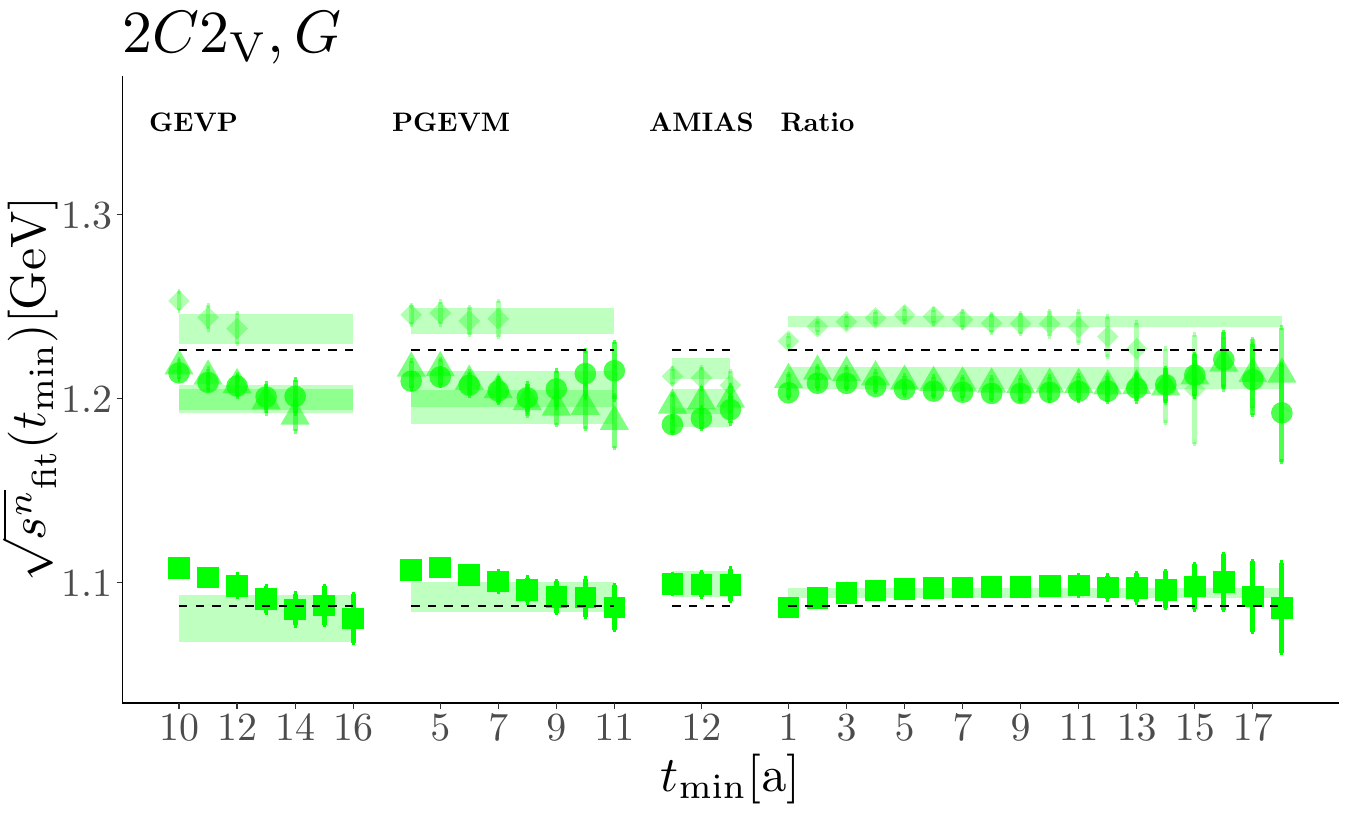}
  \caption{Same as in Fig.~\ref{fig:stability-G1u} but for irrep $2G$.}
  \label{fig:stability-2G}
\end{figure}

\begin{figure}[!h]
  \centering
  \includegraphics[width=0.9\linewidth]{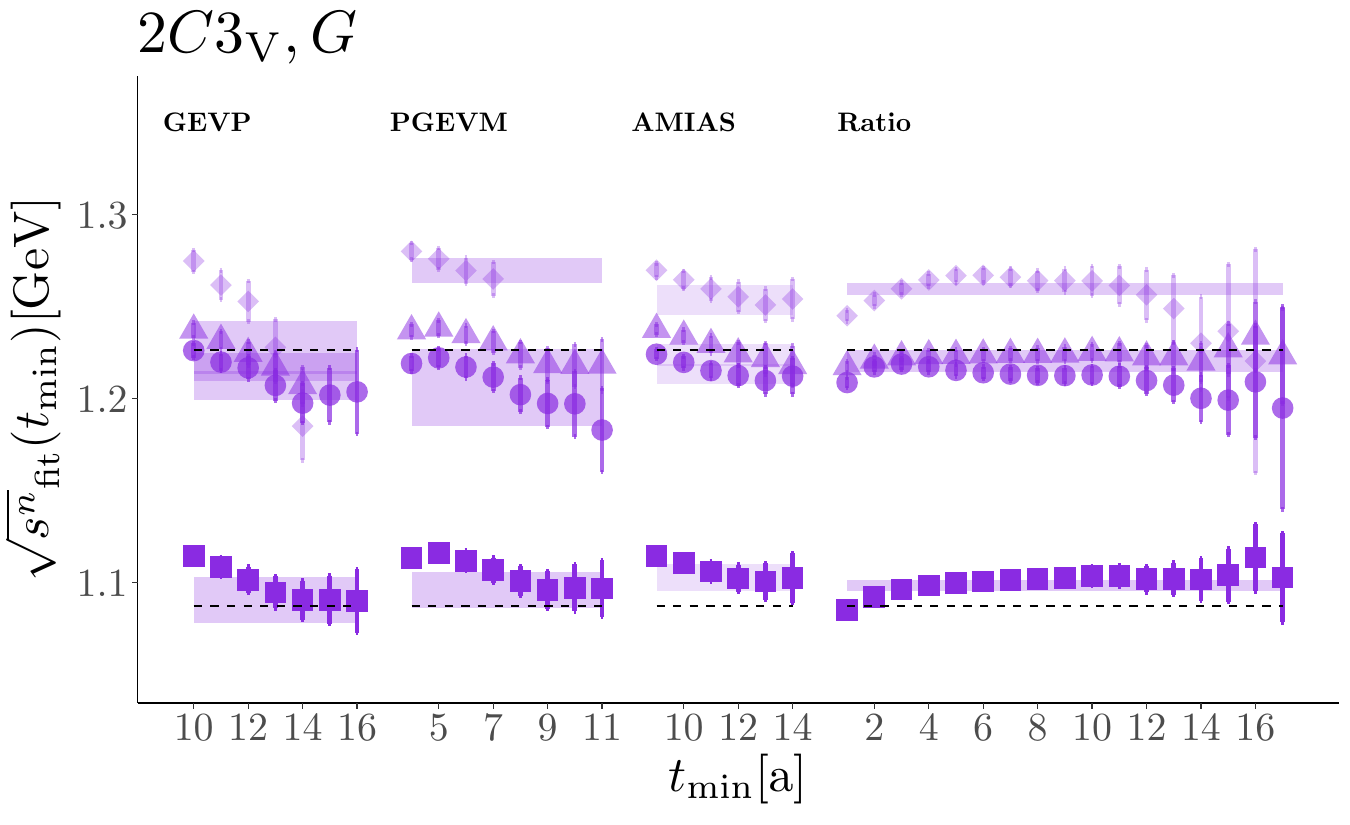}
  \caption{Same as in Fig.~\ref{fig:stability-G1u} but for irrep $3G$.}
  \label{fig:stability-3G}
\end{figure}

\begin{figure}[!h]
  \centering
  \includegraphics[width=0.9\linewidth]{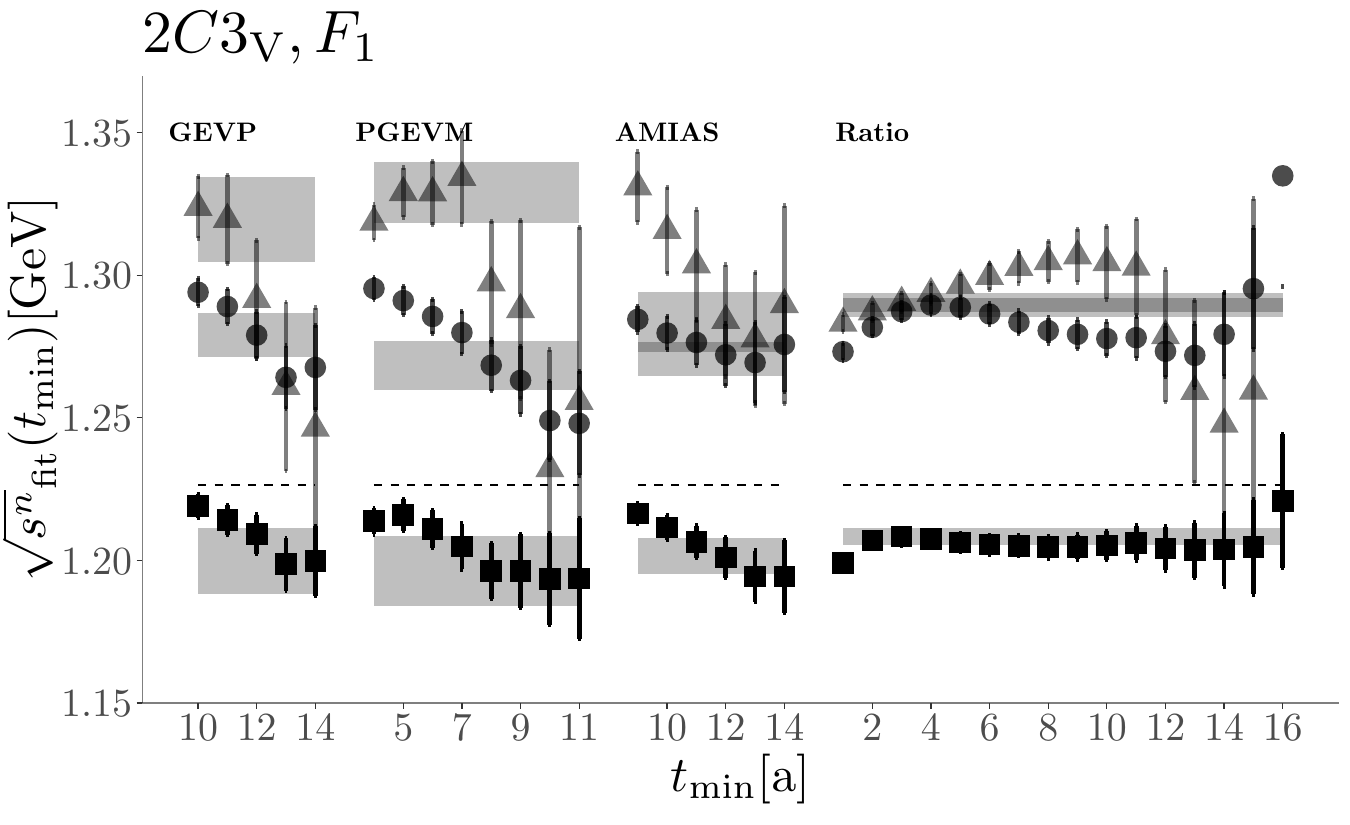} 
  \caption{Same as in Fig.~\ref{fig:stability-G1u} but for irrep $F_1$.}
  \label{fig:stability-F1}
\end{figure}

\begin{figure}[!h]
  \centering
  \includegraphics[width=0.9\linewidth]{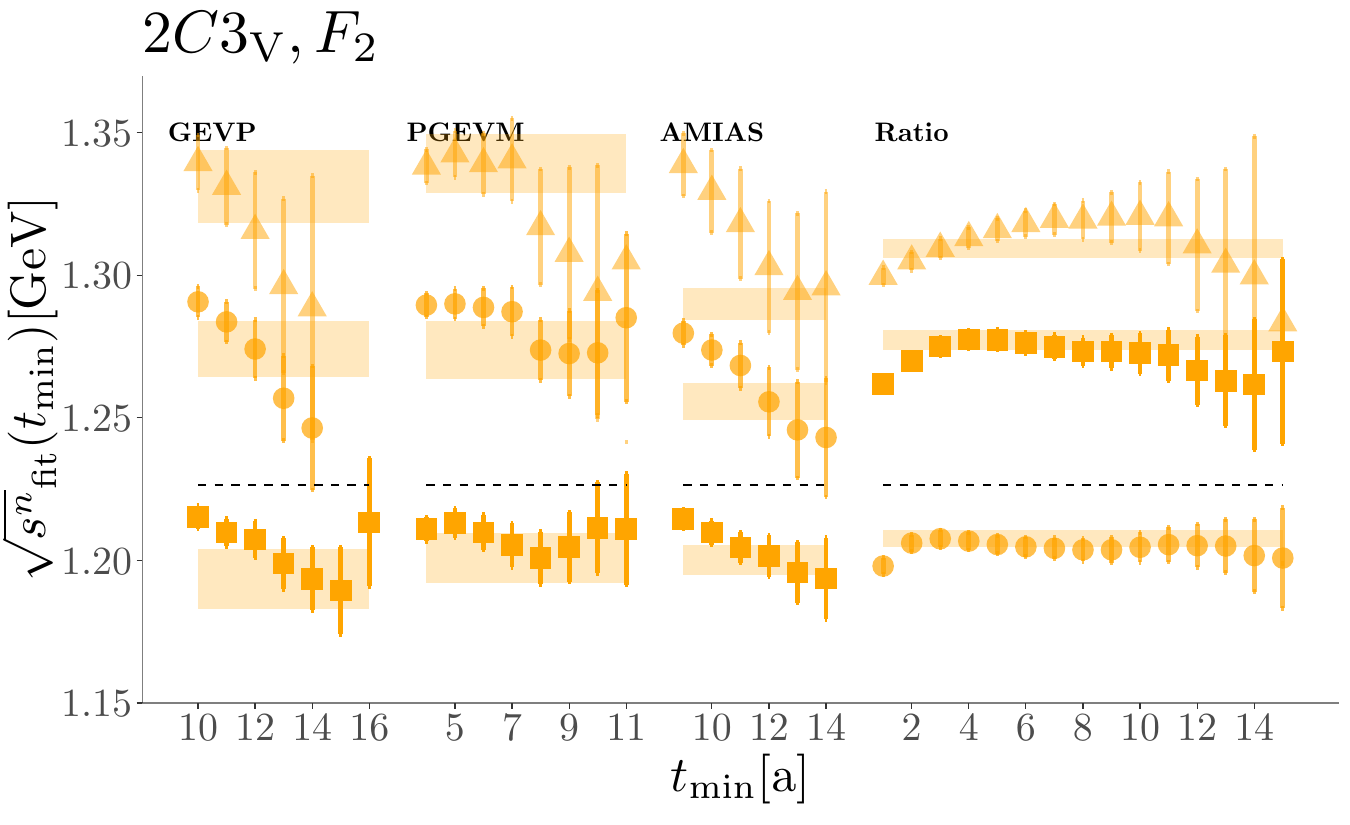}
  \caption{Same as in Fig.~\ref{fig:stability-G1u} but for irrep $F_2$.}
  \label{fig:stability-F2}
\end{figure}

In Figs.~\ref{fig:stability-G1u}, \ref{fig:stability-Hg},
\ref{fig:stability-G1}, \ref{fig:stability-G2},
\ref{fig:stability-2G}, \ref{fig:stability-3G},
\ref{fig:stability-F1}, and \ref{fig:stability-F2} we plot the
extracted energy levels in all irreps for all four methods as a
function of the lower fit-range. The figures follow the conventions of
Fig.~\ref{fig:stability-example} in the main text.

\section{Spectrum fit results}
\label{appendix:tables-results}

We list the energy levels obtained using the GEVP method, the PGEVM
method, and AMIAS in Table~\ref{tab:energy_values}. The fit range
given is the optimal one, along with the associated reduced $\chi^2$.
For GEVP and PGEVM, the quoted uncertainty of the energy levels
contains both statistical and a systematic error from the fit range
variation. For AMIAS the systematic error from varying the fit range
is negligible. For the ratio method, the error is purely
statistical.

The ratio method energy levels that were included in
Fig.~\ref{fig:spectrum_survey} are listed in
Table~\ref{tab:energy_ratio}. The quoted uncertainty is statistical
only and the closest non-interacting levels were used to obtain the
energy shift. See Sec.~\ref{sec:ratio} and
appendix~\ref{appendix:systematics} for details on how the systematic
error is computed for the ratio method, which we quote in our final
results.

\begin{table}
  \caption{ Energy levels from the $\Delta,N\pi$ correlation matrix
    for the three methods GEVP, PGEVM and AMIAS.  We use $\Pvec =
    \dvec \cdot 2\pi/L$. Shown are the cases $\dvec^2 = 0$ with group
    $O^D_h$, $\dvec^2=1$ with group $C^D_{4v}$, $\dvec^2 = 2$ with
    group $C^D_{2v}$, and $\dvec^2=3$ with $C^D_{3v}$. For each irrep
    $\Lambda$ and the lowest states labeled by $n$ we give for each
    method the optimal fit range ($t_{\rm range}=[t_{\rm min},t_{\rm
        max}]$) the associated reduced $\chi^2$ for the fit and the
    energy in the center of mass frame in lattice units.  }
  \label{tab:energy_values}
  \begin{tabular}{cccccccc}
    \hline\hline
    \multirowcell{2}{$n$}
    & \multicolumn{3}{c}{\textbf{GEVP}}
    & \multicolumn{3}{c}{\textbf{PGEVM}}
    & \textbf{AMIAS}\\
    & \makecell[c]{$t_{\rm range}$\\$[a]$} & $\frac{\chi^{2}}{\mathrm{dof}}$ & $a\sqrt{s_n^{\Lambda,\vec{P}}}$ & \makecell[c]{$t_{\rm range}$\\$[a]$}  & $\frac{\chi^{2}}{\mathrm{dof}}$ & $a\sqrt{s_n^{\Lambda,\vec{P}}}$ & $a\sqrt{s_n^{\Lambda,\vec{P}}}$ \\
    \hline
    \multicolumn{8}{c}{$\dvec^2=0$; $\Lambda = G_{1u}$}\\
    1 & [16,23] & 1.77 & 0.4375(55) & [10,23] & 1.15 & 0.4388(42) & 0.4381(29) \\
    2 & [15,20] & 0.26 & 0.5164(73) & [10,20] & 0.87 & 0.5137(61) & 0.5107(36)\\
    3 & [11,17] & 0.46 & 0.5726(43) & [7,17]  & 0.76 & 0.5732(48) & 0.5904(100)\\
    \hline
    \multicolumn{8}{c}{$\dvec^2=0$; $\Lambda = H_{g}$}\\
    1 & [14,23] & 1.02 & 0.4914(50) & [9,23] & 0.77 & 0.4982(52) & 0.4954(24)\\
    2 & [13,20] & 0.62 & 0.5298(68) & [9,20] & 1.07 & 0.5359(70) & 0.5302(53)\\
    3 & [13,21] & 1.52 & 0.5536(43) & [9,21] & 1.57 & 0.5593(44) & 0.5572(43)\\
    \hline
    \multicolumn{8}{c}{$\dvec^2=1$; $\Lambda = G_{1}$}\\
    1 & [15,24] & 2.47 & 0.4392(38) & [9,24] & 1.85 & 0.4409(30) & 0.4443(19)\\
    2 & [14,22] & 2.50 & 0.4808(33) & [8,22] & 2.15 & 0.4832(29) & 0.4860(22)\\
    3 & [13,22] & 1.08 & 0.5071(38) & [8,22] & 0.79 & 0.5072(31) & 0.5080(20)\\
    4 & [13,23] & 0.66 & 0.5155(31) & [8,23] & 0.57 & 0.5176(27) & 0.5194(22)\\
    \hline
    \multicolumn{8}{c}{$\dvec^2=1$; $\Lambda = G_{2}$}\\
    1 & [14,24] & 1.19 & 0.4861(55) & [9,24] & 1.30 & 0.4932(58) & 0.4940(32)\\
    2 & [13,26] & 1.57 & 0.5092(31) & [9,26] & 1.71 & 0.5095(34) & 0.5136(32)\\
    3 & [12,25] & 0.99 & 0.5271(34) & [9,25] & 0.79 & 0.5218(52) & 0.5211(33)\\
    \hline
    \multicolumn{8}{c}{$\dvec^2=2$; $\Lambda = G$}\\
    1 & [16,28] & 1.91 & 0.4385(51) & [9,28] & 1.34 & 0.4432(33) & 0.4461(28)\\
    2 & [13,19] & 1.85 & 0.4874(28) & [9,18] & 1.60 & 0.4851(38) & 0.4925(22)\\
    3 & [13,26] & 2.48 & 0.4866(26) & [9,25] & 1.99 & 0.4892(39) & 0.4966(24)\\
    4 & [13,25] & 2.08 & 0.5025(33) & [6,25] & 1.78 & 0.5042(29) & 0.5088(33)\\
    \hline
    \multicolumn{8}{c}{$\dvec^2=3$; $\Lambda = G$}\\
    1 & [15,24] & 2.13 & 0.4426(50) & [9,24] & 1.27 & 0.4448(39) & 0.4476(29) \\
    2 & [13,21] & 0.79 & 0.4899(32) & [9,21] & 0.62 & 0.4860(49) & 0.4924(22) \\
    3 & [13,22] & 1.93 & 0.4940(30) & [9,22] & 1.41 & 0.4945(35) & 0.4966(24) \\
    4 & [13,23] & 2.86 & 0.4983(58) & [6,23] & 2.34 & 0.5153(27) & 0.5088(33) \\
    \hline
    \multicolumn{8}{c}{$\dvec^2=3$; $\Lambda = F_1$}\\
    1 & [14,19] & 0.91 & 0.4849(44) & [9,19] & 0.67 & 0.4857(50) & 0.4878(25) \\
    2 & [12,19] & 1.00 & 0.5195(32) & [8,18] & 1.32 & 0.5148(34) & 0.5157(24) \\
    3 & [11,19] & 1.07 & 0.5366(63) & [6,19] & 1.11 & 0.5394(43) & 0.5210(43) \\
    \hline
    \multicolumn{8}{c}{$\dvec^2=3$; $\Lambda = F_3$}\\
    1 & [14,22] & 1.60 & 0.4844(42) & [8,22] & 1.37 & 0.4874(35) & 0.4871(21) \\
    2 & [12,18] & 1.98 & 0.5172(40) & [8,18] & 1.21 & 0.5170(41) & 0.5096(26) \\
    3 & [11,19] & 0.53 & 0.5403(52) & [6,19] & 0.45 & 0.5435(42) & 0.5235(22) \\\hline
  \end{tabular}
\end{table}
  \begin{widetext}
  
\begin{table}
  \caption{Energy levels in lattice units from the $\Delta,N\pi$
    correlation matrix obtained using the ratio method for all
    kinematic frames and for multiple fit-ranges (using
    the same $t_{\rm max}$). The first $t_{\rm min}$ corresponds to 
    the one shown in Fig.~\ref{fig:spectrum_survey}, the last one 
    equals the optimal fit-range for the corresponding GEVP correlator 
    and the middle one is halfway between the previous two. We used the 
    closest non-interacting level in the ratio denominator
    Eq.~(\ref{eq:ratio}). Errors are statistical only.}
  \label{tab:energy_ratio}
  \begin{center}
    \begin{tabular}{cc@{\hskip 0.4in}c@{\hskip 0.4in}c@{\hskip 0.4in}c@{\hskip 0.4in}c@{\hskip 0.4in}c@{\hskip 0.4in}c}
      \hline\hline
      \multirowcell{2}{$n$}
      & \makecell[c]{$t_{\rm range}$\\$(\lbrace t_{\rm min},\rbrace;t_{\rm max})[a]$} & 
      \multicolumn{6}{c}{$\frac{\chi^{2}}{\mathrm{dof}}(\lbrace t_{\rm min},\rbrace),a\sqrt{s_n^{\Lambda,\vec{P}}}(\lbrace t_{\rm min},\rbrace)$}\\
      \hline
      \multicolumn{8}{c}{$\dvec^2=0$; $\Lambda = G_{1u}$}\\
      1 & [7,10,14;22] & 0.34 & 0.4416(13) 
                       & 0.39 & 0.4416(16) 
                       & 0.55 & 0.4424(27)  \\
      2 & [4,8,13;17] & 0.63 & 0.5180(12)
                      & 0.33 & 0.5169(15)
                      & 0.58 & 0.5158(42) \\
      3 & [3,7,12;15] & 3.61 & 0.5752(13)
                      & 0.30 & 0.5707(19) 
                      & 0.32 & 0.5702(64) \\
      \hline
      \multicolumn{8}{c}{$\dvec^2=0$; $\Lambda = H_{g}$}\\
      1 & [5,9,14;23] & 0.52 & 0.4990(13) 
                      & 0.43 & 0.4978(20)
                      & 0.60 & 0.4957(60) \\
      2 & [4,9,13;18] & 0.48 & 0.5292(12) 
                      & 0.52 & 0.5304(24) 
                      & 0.54 & 0.5302(74)\\
      3 & [3,8,13;17] & 1.69 & 0.5576(11) 
                      & 0.52 & 0.5550(15)
                      & 0.58 & 0.5498(47) \\
      4 & [3,7,12;17] & 0.62 & 0.5696(12) 
                       & 0.79 & 0.5695(17)
                       & 0.30 & 0.5698(59)\\
      5 & [4,7,11;14] & 0.91 & 0.6013(19) 
                      & 0.79 & 0.5695(17)
                      & 0.30 & 0.5689(59)\\
      \hline
      \multicolumn{8}{c}{$\dvec^2=1$; $\Lambda = G_{1}$}\\
      1 & [4,10,15;23] & 0.10 & 0.4431(12) 
                       & 0.11 & 0.4432(17)
                       & 0.13 & 0.4434(41)\\
      2 & [4,10,14;23] & 1.67 & 0.4873(12)
                       & 0.58 & 0.4854(16)
                       & 0.85 & 0.4856(31) \\ 
      3 & [4,10,13;20] & 0.44 & 0.5098(13) 
                       & 0.55 & 0.5091(23)
                       & 0.76 & 0.5100(46)\\
      4 & [4,10,13;20] & 0.35 & 0.5178(12)
                       & 0.43 & 0.5179(18)
                       & 0.56 & 0.5172(32) \\
      5 & [5,8,13;20] & 0.41 & 0.5265(13)
                      & 0.47 & 0.5270(18)
                      & 0.63 & 0.5285(57)\\
      \hline
      \multicolumn{8}{c}{$\dvec^2=1$; $\Lambda = G_{2}$}\\
      1 & [4,8,14;23] & 1.06 & 0.5022(13) 
                      & 0.42 & 0.4993(19)
                      & 0.41 & 0.4920(76)\\
      2 & [4,8,13;23] & 0.31 & 0.5140(12)
                      & 0.33 & 0.5135(15)
                      & 0.46 & 0.5126(32) \\
      3 & [3,7,12;18] & 0.35 & 0.5283(12) 
                      & 0.41 & 0.5279(15)
                      & 0.37 & 0.5246(38)\\
      4 & [3,7,11;21] & 1.07 & 0.5398(11)
                      & 0.46 & 0.5385(13)
                      & 0.58 & 0.5379(19) \\
      5 & [3,7,11;18] & 0.30 & 0.5479(12) 
                      & 0.31 & 0.5475(16)
                      & 0.39 & 0.5488(30)\\
      \hline
      \multicolumn{8}{c}{$\dvec^2=2$; $\Lambda = G$}\\
      1 & [3,10,16;24] & 0.81 & 0.4442(12) 
                       & 0.16 & 0.4458(19)
                       & 0.22 & 0.4466(62)\\
      2 & [3,7,13;21] & 2.02 & 0.4904(12)
                      & 0.39 & 0.4885(13)
                      & 0.56 & 0.4894(29) \\
      3 & [3,7,13;22] & 3.44 & 0.4928(12)
                      & 0.26 & 0.4905(13)
                      & 0.37 & 0.4899(23)\\
      4 & [3,6,12;17] & 1.53 & 0.5040(12)
                      & 0.58 & 0.5052(15)
                      & 0.49 & 0.5007(44)\\
      5 & [3,6,11;17] & 1.52 & 0.5154(12)
                      & 0.66 & 0.5157(14)
                      & 0.41 & 0.5127(28)\\
      \hline
      \multicolumn{8}{c}{$\dvec^2=3$; $\Lambda = G$}\\
      1 & [4,10,15;21] & 0.61 & 0.4460(12)
                       & 0.17 & 0.4480(20)
                       & 0.25 & 0.4481(58)\\
      2 & [4,10,15;21] & 1.10 & 0.4941(13) 
                       & 0.24 & 0.4923(20)
                       & 0.14 & 0.4867(76)\\
      3 & [4,8,13;23] & 0.24 & 0.4965(12)
                      & 0.28 & 0.4968(15) 
                      & 0.34 & 0.4963(32) \\
      4 & [3,8,13;18] & 3.75 & 0.5113(13) 
                      & 0.30 & 0.5131(22)
                      & 0.34 & 0.5069(74)\\
      5 & [3,8,13;18] & 0.12 & 0.5213(12) 
                      & 0.13 & 0.5213(15)
                      & 0.05 & 0.5192(30)\\
      \hline
      \multicolumn{8}{c}{$\dvec^2=3$; $\Lambda = F_1$}\\
      1 & [3,10,14;20] & 0.65 & 0.4905(12) 
                       & 0.27 & 0.4892(20)
                       & 0.40 & 0.4886(52)\\
      2 & [5,9,12;18] & 1.54 & 0.5231(14) 
                      & 0.49 & 0.5198(18)
                      & 0.54 & 0.5169(37) \\
      3 & [3,7,11;17] & 2.06 & 0.5238(13) 
                      & 0.55 & 0.5288(23)
                      & 0.80 & 0.5288(69)\\
      \hline
      \multicolumn{8}{c}{$\dvec^2=3$; $\Lambda = F_2$}\\ 
      1 & [3,10,14;19] & 0.72 & 0.4902(12) 
                       & 0.34 & 0.4890(21) 
                       & 0.60 & 0.4878(52) \\
      2 & [5,8,12;19]  & 0.46 & 0.5184(14) 
                       & 0.27 & 0.5167(20)
                       & 0.30 & 0.5140(49) \\
      3 & [3,7,11;17] & 1.68 & 0.5315(14) 
                      & 0.11 & 0.5356(22)
                      & 0.17 & 0.5354(69)\\
      \hline
    \end{tabular}
  \end{center}
\end{table}
\end{widetext}

\section{Optimization of quantiziation condition solution}
\label{appendix:optimization}
As explained in the text, the statistical errors quoted for the
resonance parameters are obtained via jackknife resampling.  To
determine these, we perform the minimization of $\chi^2$ in each
jackknife sample, which in turn requires evaluating the so-called zeta
function and finding the roots of the quantization condition in every
iteration of the $\chi^2$ minimization algorithm for every jackknife
sample. While this is carried out explicitly for our final results
that use the ratio method, for the fits to the spectrum results using
the GEVP, PGEVM, and AMIAS methods shown in
Table~\ref{tab:qc_fit_systematics_test}, we accelerate this process by
employing a multi-dimensional spline interpolation in the parameters
being fitted and evaluate the quantization condition at the grid
points of this interpolation. The spline interpolation function is
uniquely defined by the values of each finite volume energy level at
each grid point $E^{\vec{P},\Lambda}(M_{R,l},\Lambda_{R,m},a_{0,n})$
with $l = 0,\cdots,L-1,\,\, m = 0 ,\cdots, M-1$ and $n = 0, \cdots,
N-1$.  We determine our grid by first carrying out trial fits of our
data, obtaining the errors of the fit parameters via their covariance
matrix. From these initial fits, we found that the following grid
contains the parameters.
\begin{align}
  M_{R,i}      \, &\in \, [1100 \,,\, 1500] \, {\mathrm{MeV}} \,,
  \nonumber\\
  \Gamma_{R,j} \, &\in \, [50\,,\,300]      \, {\mathrm{MeV}} \,,
  \nonumber\\ 
  a_{0,k}      \, &\in \, [-0.0025\, ,\, -0.0001] \, {\mathrm{MeV}}^{-1}\,.
\end{align} 
The number of points in each dimension is taken as $L=50$, $M=50$, and $N=8$, which we ascertain is fine enough to reproduce all our trial fit results.
The definition of our spline interpolation function can be written as:
\begin{align}
  S(M_R,\Gamma_R,a_0)=\sum_{i=0}^{3}\sum_{j=0}^{3}\sum_{k=0}^{3}\,\, C^{i,j,k}_{l,m,n}\, {t}_{l}^i \, {u}_{m}^j \, {r}_n^k\,,
  \label{eq:spline}
\end{align}
where $(M_R,\, \Gamma_R, \, a_0) \, \in \, [M_{R,l},\, M_{R,l+1}]\times [\Gamma_{R,n},\, \Gamma_{R,n+1}]\times [a_{0,l},\, a_{0,l+1}]$,
and $t,\,u,\,r$ are the following dimensionless coefficients
\begin{align}
  t_{l} &= \frac{M_R-M_{R,l}}{\Delta M_{R,l}}\,,
  \nonumber\\
  u_{m} &= \frac{\Gamma_R-\Gamma_{R,m}}{\Delta \Gamma_{R,m}}\,,
  \nonumber\\
  r_{n} &= \frac{a_0-a_{0,n}}{\Delta a_{0,n}}\,,
\end{align}
with the corresponding widths
\begin{align}
  \Delta M_{R,l} &=   M_{R,l+1}-M_{R,l}\,,
  \nonumber\\
  \Delta \Gamma_{R,m} &= \Gamma_{R,m+1}-\Gamma_{R,m}\,,
  \nonumber\\
  \Delta a_{0,n} &= a_{0,n+1}-a_{0,n}\,.
\end{align}
The spline coefficients $C$ in Eq. \refeq{eq:spline} are determined using the algorithm detailed in Ref.~\cite{Endrodi:2010ai}. 

\section{Systematic uncertainties in the ratio method}
\label{appendix:systematics}
For our final results for the resonance mass and resonance width, we
use the results from the ratio method and obtain a systematic by
varying the fit to the L\"uscher quantization as described in
Sec.~\ref{sec:ratio}. This yields 45 fits, the results of which are
collected in Table~\ref{tab:qc_fit_systematics}.

\begin{table}[h]
  \caption{ Results for the scattering parameters, namely the
    resonance mass, $M_R$, resonance width, $\Gamma_R$, and scattering
    length, $M_\pi a_0$, using the L\"uscher quantization condition
    and energy levels determined from the ratio method. We delineate
    two of the five fit variations listed in Sec.~\ref{sec:ratio} via
    the table sub-headings, namely the variation arising from 1.)
    varying the number of energy levels included and 2.) whether
    considering only $P$- or $P$- and $S$-wave. For 3.), we give the
    value of the scattering length $a_0$ when this is left as a free
    parameter and omit it when it is fixed via the direct evaluation.
    The first column labels the remaining two variations, namely 4.)
    that obtained by varying the fit-range in three ways (first index)
    and 5.) by varying the back-to-back momenta of the non-interacting
    pion-nucleon operator used in the denominator of the ratio method
    (second index).
    \label{tab:qc_fit_systematics}}
  \begin{center}
    \begin{tabular}{cr@{(}lr@{(}lr@{.}lr@{.}l}\hline\hline
      \multirow{2}{*}{Fit type} & \multicolumn{4}{c}{Breit-Wigner parameters}                                                    & \multicolumn{2}{c}{\multirow{2}{*}{$M_\pi a_0$}} & \multicolumn{2}{c}{\multirow{2}{*}{$\chi^2/$dof}                        }  \\
                                & \multicolumn{2}{c}{$M_R$ [MeV]}             & \multicolumn{2}{c}{$\Gamma_R$ [MeV]}             & \multicolumn{2}{c}{\quad}                        & \multicolumn{2}{c}{\quad}\\\hline
      \multicolumn{9}{c}{($J,\ell$) = (1/2, 0) and (3/2, 1); No. of $\sqrt{s}$ points = 14}                                                                                                            \\
      1,1                       & 1287                                        & 7)                                               & 98                        & 19)  & -0                    & 12(5)  & 1 & 34 \\
      1,2                       & 1289                                        & 17)                                              & 169                       & 72)  & -0                    & 16(7)  & 0 & 37 \\  
      1,3                       & 1291                                        & 20)                                              & 241                       & 4)   & -0                    & 17(17) & 0 & 21 \\
      2,1                       & 1279                                        & 5)                                               & 83                        & 15)  & -0                    & 12(5)  & 1 & 74 \\
      2,2                       & 1280                                        & 14)                                              & 140                       & 54)  & -0                    & 16(7)  & 0 & 47 \\
      2,3                       & 1282                                        & 52)                                              & 193                       & 250) & -0                    & 17(17) & 0 & 22 \\
      3,1                       & 1272                                        & 5)                                               & 70                        & 12)  & -0                    & 12(5)  & 2 & 18 \\
      3,2                       & 1274                                        & 13)                                              & 121                       & 45)  & -0                    & 16(7)  & 0 & 54 \\
      3,3                       & 1289                                        & 60)                                              & 207                       & 288) & -0                    & 17(3)  & 0 & 22 \\
      1,1                       & 1287                                        & 7)                                               & 98                        & 19)  & \multicolumn{2}{c}{-} & 1      & 24     \\
      1,2                       & 1289                                        & 16)                                              & 161                       & 66)  & \multicolumn{2}{c}{-} & 0      & 40     \\  
      1,3                       & 1286                                        & 52)                                              & 182                       & 221) & \multicolumn{2}{c}{-} & 0      & 27     \\
      2,1                       & 1279                                        & 5)                                               & 83                        & 15)  & \multicolumn{2}{c}{-} & 1      & 61     \\
      2,2                       & 1280                                        & 14)                                              & 134                       & 40)  & \multicolumn{2}{c}{-} & 2      & 80     \\
      2,3                       & 1278                                        & 43)                                              & 150                       & 166) & \multicolumn{2}{c}{-} & 0      & 28     \\
      3,1                       & 1272                                        & 5)                                               & 70                        & 12)  & \multicolumn{2}{c}{-} & 2      & 01     \\
      3,2                       & 1274                                        & 6 )                                              & 117                       & 20)  & \multicolumn{2}{c}{-} & 0      & 54     \\
      3,3                       & 1285                                        & 50)                                              & 163                       & 195) & \multicolumn{2}{c}{-} & 0      & 26     \\
      \multicolumn{9}{c}{\quad}\\\hline
      \multicolumn{9}{c}{($J,\ell$) = (1/2, 0) and (3/2, 1); No. of $\sqrt{s}$ points = 12}                                                                                                            \\
      1,1                       & 1249                                        & 10)                                              & 26                        & 17)  & -0                    & 11(5)  & 0 & 98 \\
      1,2                       & 1258                                        & 27)                                              & 72                        & 72)  & -0                    & 15(7)  & 0 & 30 \\  
      1,3                       & 1221                                        & 22)                                              & 17                        & 33)  & -0                    & 13(18) & 0 & 14 \\
      2,1                       & 1246                                        & 10)                                              & 23                        & 15)  & -0                    & 11(5)  & 1 & 26 \\
      2,2                       & 1252                                        & 23)                                              & 60                        & 56)  & -0                    & 15(7)  & 0 & 35 \\
      2,3                       & 1220                                        & 21)                                              & 16                        & 30)  & -0                    & 14(19) & 0 & 13 \\
      3,1                       & 1243                                        & 8)                                               & 22                        & 12)  & -0                    & 11(5)  & 1 & 60 \\
      3,2                       & 1250                                        & 21)                                              & 57                        & 51)  & -0                    & 15(7)  & 0 & 39 \\
      3,3                       & 1224                                        & 26)                                              & 21                        & 41)  & -0                    & 13(18) & 0 & 17 \\
      1,1                       & 1250                                        & 10)                                              & 27                        & 17)  & \multicolumn{2}{c}{-} & 0      & 89     \\
      1,2                       & 1257                                        & 26)                                              & 68                        & 67)  & \multicolumn{2}{c}{-} & 0      & 28     \\  
      1,3                       & 1221                                        & 22)                                              & 17                        & 33)  & \multicolumn{2}{c}{-} & 0      & 13     \\
      2,1                       & 1246                                        & 9)                                               & 24                        & 15)  & \multicolumn{2}{c}{-} & 1      & 15     \\
      2,2                       & 1251                                        & 21)                                              & 58                        & 52)  & \multicolumn{2}{c}{-} & 0      & 32     \\
      2,3                       & 1220                                        & 22)                                              & 16                        & 30)  & \multicolumn{2}{c}{-} & 0      & 12     \\
      3,1                       & 1243                                        & 8)                                               & 22                        & 13)  & \multicolumn{2}{c}{-} & 1      & 45     \\
      3,2                       & 1249                                        & 20)                                              & 54                        & 48)  & \multicolumn{2}{c}{-} & 0      & 36     \\
      3,3                       & 1224                                        & 27)                                              & 21                        & 40)  & \multicolumn{2}{c}{-} & 0      & 15     \\
      \multicolumn{9}{c}{\quad}\\\hline
      \multicolumn{9}{c}{($J,\ell$) = (3/2, 1); No. of $\sqrt{s}$ points = 5}                                                                                                                          \\
      1,1                       & 1282                                        & 7)                                               & 113                       & 22)  & \multicolumn{2}{c}{-} & 0      & 77     \\
      1,2                       & 1284                                        & 18)                                              & 180                       & 87)  & \multicolumn{2}{c}{-} & 0      & 45     \\  
      1,3                       & 1265                                        & 60)                                              & 227                       & 355) & \multicolumn{2}{c}{-} & 0      & 17     \\
      2,1                       & 1272                                        & 6)                                               & 93                        & 16)  & \multicolumn{2}{c}{-} & 1      & 17     \\
      2,2                       & 1273                                        & 14)                                              & 145                       & 58)  & \multicolumn{2}{c}{-} & 0      & 45     \\
      2,3                       & 1253                                        & 44)                                              & 167                       & 225) & \multicolumn{2}{c}{-} & 0      & 15     \\
      3,1                       & 1264                                        & 5)                                               & 77                        & 13)  & \multicolumn{2}{c}{-} & 1      & 76     \\
      3,2                       & 1265                                        & 13)                                              & 122                       & 48)  & \multicolumn{2}{c}{-} & 0      & 11     \\
      3,3                       & 1261                                        & 54)                                              & 198                       & 296) & \multicolumn{2}{c}{-} & 0      & 20     \\
                                                                                                                                                                                                              \\    \end{tabular}      \end{center}
\end{table}

\end{document}